\documentclass[journal, a4paper]{IEEEtran}
\IEEEoverridecommandlockouts
\usepackage{cite}
\usepackage{amsmath,amssymb,amsfonts,mathtools}
\usepackage{bm}
\usepackage{bbm}
\usepackage{graphicx}
\usepackage{textcomp}
\usepackage{xcolor}
\usepackage{subcaption}
\usepackage{multirow}
\usepackage{siunitx}
\usepackage{algorithm}
\usepackage{algpseudocode}
\usepackage{cleveref}
\usepackage{url}
\usepackage{babel}
\DeclareSIUnit{\belmilliwatt}{Bm}
\DeclareSIUnit{\dBm}{\deci\belmilliwatt}

\def\BibTeX{{\rm B\kern-.05em{\sc i\kern-.025em b}\kern-.08em
    T\kern-.1667em\lower.7ex\hbox{E}\kern-.125emX}}
    
\definecolor{rick}{RGB}{0,139,0}

\definecolor{Rafael}{RGB}{139,0,0}

\definecolor{Muah}{RGB}{0,0,139}

\begin{document}

\title{Diffusion Models for Accurate Channel Distribution Generation}

\author{Muah Kim,~\IEEEmembership{Student Member,~IEEE}, 
    Rick Fritschek~\IEEEmembership{Member,~IEEE}, 
    and Rafael F. Schaefer,~\IEEEmembership{Senior Member,~IEEE}

\thanks{This work was supported in part by the German Federal Ministry of Education and Research (BMBF) within the National Initiative on 6G Communication Systems through the Research Hub \emph{6G-life} under Grant 16KISK001K, in part by the German Research Foundation (DFG) as part of Germany’s Excellence Strategy – EXC 2050/1 – Project ID 390696704 – Cluster of Excellence \emph{``Centre for Tactile Internet with Human-in-the-Loop'' (CeTI)} of Technische Universit\"at Dresden, and in part by the DFG under Grant SCHA 1944/7-1. An earlier version of this paper was presented in part at the \textit{ITG International Workshop on Smart Antennas and Conference on Systems, Communications, and Coding}, Braunschweig, Germany, Feb. 2023 \cite{Kim2023diff}.}
\thanks{
M. Kim and R. Fritschek are with the Chair of Information Theory and Machine Learning, Technische Universit\"at Dresden, 01062 Dresden, Germany. (e-mail:~\{muah.kim, rick.fritschek\}@tu-dresden.de)
}
\thanks{
R. F. Schaefer is with the Chair of Information Theory and Machine Learning, the BMBF Research Hub 6G-life, the Cluster of Excellence ``Centre for Tactile Internet with Human-in-the-Loop (CeTI),'' and the 5G Lab Germany, Technische Universit\"at Dresden, 01062 Dresden, Germany (e-mail: rafael.schaefer@tu-dresden.de).}
}


\maketitle

\begin{abstract}
Strong generative models can accurately learn channel distributions. This could save recurring costs for physical measurements of the channel. Moreover, the resulting differentiable channel model supports training neural encoders by enabling gradient-based optimization.
The initial approach in the literature draws upon the modern advancements in image generation, utilizing generative adversarial networks (GANs) or their enhanced variants to generate channel distributions. In this paper, we address this channel approximation challenge with diffusion models (DMs), which have demonstrated high sample quality and mode coverage in image generation. In addition to testing the generative performance of the channel distributions, we use an end-to-end (E2E) coded-modulation framework underpinned by DMs and propose an efficient training algorithm. 
Our simulations with various channel models show that a DM can accurately learn
channel distributions, enabling an E2E framework to achieve near-optimal symbol error rates (SERs). 
Furthermore, we examine the trade-off between mode coverage and sampling speed through skipped sampling using sliced Wasserstein distance (SWD) and the E2E SER. We investigate the effect of noise scheduling on this trade-off, demonstrating that with an appropriate choice of parameters and techniques, sampling time can be significantly reduced with a minor increase in SWD and SER. Finally, we show that the DM can generate a correlated fading channel, whereas a strong GAN variant fails to learn the covariance. 
This paper highlights the potential benefits of using DMs for learning channel distributions, which could be further investigated for various channels and advanced techniques of DMs.

\end{abstract}


\section{Introduction}
\subsection{Learning Channel Distributions by Deep Learning}
Communication engineering has dedicated itself to establishing good mathematical channel models and developing optimal communication techniques based on those models. Despite the remarkable progress during the past century, such an approach has limitations in some scenarios, for example, when real channels are essentially not easy to mathematically describe or optimization problems are hard to solve even with a perfect channel model. To push the limit further, machine learning has been actively utilized for channel modeling \cite{wang2020data, xia2022generative, xiao2022channelgan} and estimation \cite{GAN-ChEst2021, arvinte2022score, zicheng2021deep, balevi2020high, weisser2021generative, orekondy2022mimo, fesl2024diffusion}.

Another approach is to learn the channel distribution as a black box having a conditional distribution by using generative networks. The straightforward benefit of the learned channel is to reproduce the channel coefficients without repeating physical experiments if the channel is stationary. Another application of the learned channel is to support machine learning-based optimization of E2E channel coding and or modulation frameworks. 
Recent works aim to leverage deep learning's strong performance for classification tasks to design channel codes and modulation schemes that come close to or surpass classical schemes \cite{park2021high, jiang2020joint, sattiraju2018performance, ye2019circular, gautam2020blind, lim2022hybrid, abdelmoniem2019enhanced, balevi2020autoencoder}, and they require differentiable channel models to enable backpropagation from the receiver to the transmitter through the channel. When an accurate and differentiable channel model is not available or does not exist, the channel block can be substituted by a generative neural network (NN) that is differentiable and provides a similar distribution. The idea of learning channel distributions using generative models was first proposed in \cite{ye2018channel}. This approach was refined and expanded using advanced variants of generative adversarial networks (GANs) \cite{dorner2020wgan, GAN-OShea, o2019approximating}. For the same reason, DMs have been used in communication engineering \cite{}. 

Carefully designed GANs have shown great performance for simple channels in the literature, but they might face limitations with more complex channels, as shown in \cite{dorner2020wgan} for a tapped delay line channel. The important metric for learning channel distributions is mode coverage, which is the ability to capture the sample distribution, rather than sample quality. It is well known that GANs can generate quality samples but suffer from poor mode coverage, known as mode collapse because their training goal is to generate realistic samples that can effectively fool the discriminator, which does not necessarily learn the entire sample distribution. This motivated us to tackle the same problem with diffusion models (DMs), as they have been shown to provide better mode coverage in computer vision.

DMs generate data samples by adding noise until the samples follow a normal distribution, and then learn to denoise these noisy samples to reconstruct the original data. By applying the denoising procedure to samples from the normal distribution, new samples can be generated. While DMs might take several forms, there are three principal variants characterized by unique methodologies for perturbing training samples and learning to denoise them \cite{croitoru2022diffusion}. These include diffusion-denoising probabilistic model (DDPM) \cite{sohl2015deep, ho2020denoising}, noise-conditioned score networks (NCSNs) \cite{song2019generative}, and models based on stochastic differential equations (SDEs) \cite{song2020score}. This study uses a DDPM to generate a channel distribution, a decision motivated by a substantial body of research utilizing DDPMs for conditional generation. Typically, channels are defined by a conditional distribution, making DDPMs an apt choice for our purpose.

\subsection{Related Works}
Several studies that closely align with our work have explored learning channel distributions with GANs.
This approach was first studied in \cite{ye2018channel}, where a conditional GAN was evaluated using the symbol error rate (SER) of the NN-based E2E communication framework for an additive white Gaussian noise (AWGN) channel and a Rayleigh fading channel. Comparisons were made to traditional channel coding and modulation methods, as well as to an AE trained with coherent detection and estimated channels with pilot signals. This conditional GAN-based E2E framework achieved the same SER as a Hamming code with a quadrature amplitude modulation (QAM).

This approach saw further advancements in \cite{dorner2020wgan}, where a Wasserstein GAN (WGAN) was utilized for robust channel estimation performance over an over-the-air channel. Comparisons were made to a reinforcement learning (RL)-based optimization of the neural encoder and other frameworks such as an AE trained with a simulated tapped delay line (TDL) channel model and a QAM with neural decoder. The results suggest that using the WGAN can provide more accurate results than using a non-adaptive channel model or standard modulation like QAMs, while requiring less computational complexity compared to RL and achieving the same bit error rate (BER).  
However, the WGAN encountered mode collapse when trained for a TDL model.

\subsection{Main Contributions}
Image generation and channel approximation represent two distinct application areas, where the optimal solution in one does not necessarily translate to superior performance in the other. 
This paper aims to establish a baseline of DM-based channel generation with a simple framework and explore how certain design parameters relate to specific performance metrics. Upon this work, extensions can be made for more sophisticated and specific communication scenarios.
The specific contributions are as follows:
\begin{itemize}
    \item We present a DM design for learning channels that is compatible with the E2E NN-based communication framework in the literature. 
    For the E2E frameworks, we employ an efficient pre-training algorithm. This not only provides faster and more stable convergence but also achieves smaller E2E SER compared to the commonly used iterative training algorithm. 
    \item This paper demonstrates a DM's advanced generative performance through channel models with additive noise, random fading, non-linear power amplification, and correlated fading. Compared to the exact channel models, the DMs have shown negligible differences, outperforming WGANs in most cases.
    \item To mitigate the slow sampling time of DMs, we implemented skipped sampling and the denoising diffusion implicit model (DDIM) \cite{song2021denoising}, which is known to work well with skipped sampling. To minimize the performance degradation from skipped sampling for channel generation, we investigated the effects of noise scheduling and parametrization techniques \cite{lin2023common, salimans2022progressive} using the sliced Wasserstein distance (SWD) and the E2E SER. Simulation results show that a good choice of these parameters can improve the trade-off between generative performance and sampling time.
\end{itemize}

\subsection{Notation}
A stochastic process $\bm{x}_1, \bm{x}_2, \dots, \bm{x}_t$ indexed by time step $t$ can be shortened as $\bm{x}_{1:t}$. An $n$-dimensional random vector is represented as $\bm{X}^n$, and $X_i$ refers to its $i^{\text{th}}$ element for $i=1,2,\dots, n$. The $n$-dimensional identity matrix is denoted by $\mathbf{I}_n$. The $\ell_2$-norm of a vector $\bm{x}$ is denoted by $||\bm{x}||_2$. When a probability distribution, function, scalar, vector, or matrix has the subscript $\theta$, it means they are parametric and can be learned by NNs. In this paper, $\mathcal{U}\{a, b\}$ indicates the discrete uniform distribution over the integers in $[a,b]$ for arbitrary integers $a,b$ such that $a<b$. When we refer to the distribution that a normal random variable $X$ follows, the notation $X\sim\mathcal{N}(\mu, \sigma^2)$ is used for $\mu\in\mathbb{R}$ and $\sigma>0$. When referring to the probability distribution $p(X)$ of a random variable $X$, we use the notation $p(X)=\mathcal{N}(X;\mu, \sigma^2)$, where the variable and parameters are separated by a semicolon.

\section{End-to-end Framework and Algorithms}
This section defines the E2E framework that utilizes the generated channel block. The framework consists of an NN-based encoder and decoder, which are connected by the emulated channel block. Subsequently, we explain two training algorithms for this E2E framework, along with their respective advantages and disadvantages. 

\subsection{E2E Coded Modulation Framework}\label{ssec:AE_cDiff}
\begin{figure}[t!]
    \begin{subfigure}{\columnwidth}
        \centering
        \includegraphics[width=\columnwidth]{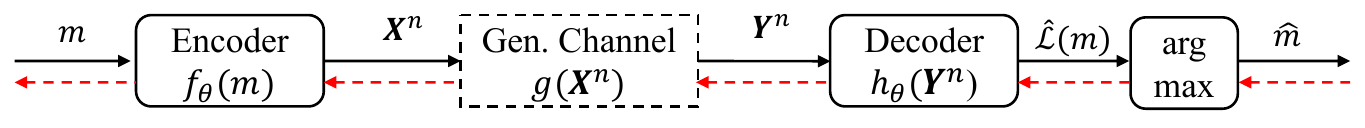}
        \subcaption{Training the encoder by using the generated channel. The dashed red arrows show the backpropagation path.}\label{fig:block_diagram_AE}
    \end{subfigure}
    \centering
    \begin{subfigure}{\columnwidth}
        \centering
        \vspace{0.2cm}
        \includegraphics[width=0.98\columnwidth]{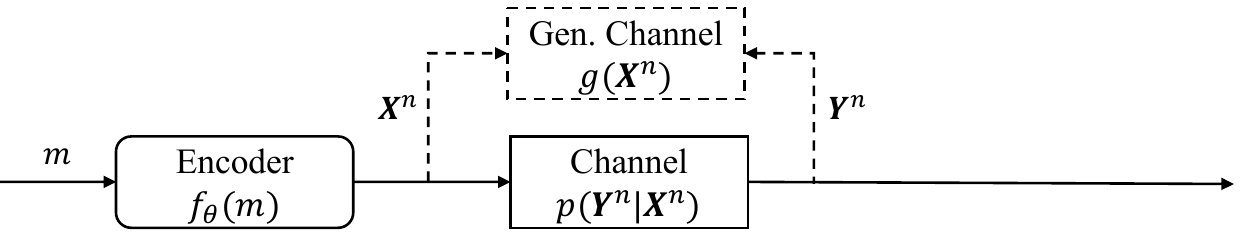}
        \subcaption{Encoder-specific training of the generative model.} \label{fig:block_diagram_gen}
    \end{subfigure}  
    \centering
    \begin{subfigure}{\columnwidth}
        \centering
        \vspace{0.2cm}
        \includegraphics[width=0.98\columnwidth]{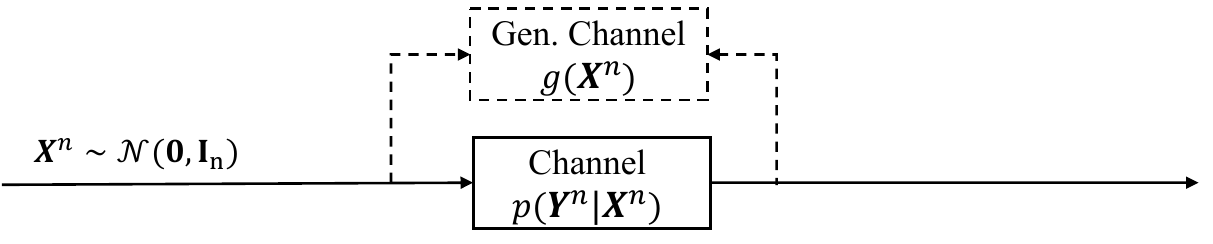}
        \subcaption{Training the generative model with random channel input.} \label{fig:block_diagram_gen_PreT}
    \end{subfigure}  
    \caption{Block diagrams of the E2E framework.}\label{fig:block_diagram}
\end{figure}
Consider an AE model where a generated channel is embedded between the encoder and the decoder blocks, replacing the real channel as depicted in Fig. \ref{fig:block_diagram_AE}. The neural encoder $f_{\theta}$ takes the message $m\in\{1,2,\dots,M\}$, converts it to a one-hot encoded message, and then passes it through an NN, outputting a codeword in an $n$-dimensional real vector space, i.e., $f_{\theta}(m)\in\mathbb{R}^n$. We use an NN whose input and output layers have $M$ and $n$ nodes, respectively, with several hidden layers in between. The encoder model's output is normalized to prevent diverging and to satisfy the power constraint. We denote the generated channel output as $g(f_{\theta}(m))\in\mathbb{R}^n$, which is sampled from the generative model.
The neural decoder $h_{\theta}$ mirrors the encoder's structure, taking $n$-dimensional real vectors and producing $M$ values indicating the predicted likelihood of each message, i.e., $\hat{\mathcal{L}}(m)=h_{\theta}\left( g(f_{\theta}(m)) \right) \in (\mathbb{R}^+)^M$. Then, the argument giving the maximum likelihood is taken to be the decoded symbol $\hat{m}$.
Note that this framework is not particularly restricted to either a joint channel coding and modulation or a coded modulation, as the input message $m$ can be either uncoded sequences or channel-encoded sequences. However, we use the term coded modulation in this work. 

It is worth noting that the decoder can be trained independently of the channel's differentiability. However, training the encoder necessitates a differentiable channel model. To meet this requirement, a generated channel is considered, which is differentiable (and can form a backpropagation path) as shown in Fig. \ref{fig:block_diagram_AE}. The AE model is trained to minimize the cross entropy of the one-hot-encoded message and the decoder output, i.e.,
\begin{align}
\mathcal{L}_{AE}\! =\! \mathbb{E}_{m\sim p(m)}\left[ -\sum_{i=1}^M\mathbbm{1}_{i=m}\log \left( h_\theta\left(g\left(f_\theta(m)\right)\right) \right)_i \right],
\end{align}
where $\mathbbm{1}$ is the indicator function. 
In the next subsection, we outline the alternation between training the AE and the generative model.

\subsection{E2E Training Algorithms}\label{ssec:training_alg}
We consider two E2E training algorithms: the \textit{iterative-} and the \textit{pre-training algorithms}. The former trains the generative model and the AE iteratively, while the latter extensively trains the generative model with random channel inputs first and then optimizes the AE. The iterative training algorithm is commonly used in the literature for learning E2E coded modulation with a generative NN \cite{GAN-OShea, dorner2020wgan, Kim2023diff}.

The iterative training algorithm has two training phases and alternates between them until the models converge. 
In the first phase, the generative model is trained with the encoder-specific dataset, comprised of a set of randomly sampled $M$ encoded channel inputs, following the message distribution, and the corresponding channel outputs, as depicted in Fig. \ref{fig:block_diagram_gen}. 
In the second phase, the AE is trained with the backpropagation path made by the generated channel outputs. 
These two phases are repeated until the AE converges and the E2E SER reaches an acceptable level. When the generative network is trained with the encoder-specific dataset, the generative model only needs to learn $M$ different channel output distributions $p(\bm{Y}^n|\bm{X}^n=f_\theta(m))$ for $m\in\{1,2,\dots,M\}$, simplifying the task compared to learning the channel output distribution for arbitrary channel inputs. While this may simplify the learning task, this iterative manner can lead to extensive training time and unstable convergence.
Since the generative model is encoder-specific, frequent updates are required when changes occur in the encoder. Otherwise, the generated channel output can significantly deviate from the ground truth, misguiding the encoder. The utility of the generative model is also limited, as it is only compatible with the encoder used during training. 

Given these shortcomings, we propose a pre-training algorithm, where the generative model is thoroughly trained once with arbitrary channel inputs and their channel output distributions. Specifically, we use the isotropic standard normal distribution to generate random channel inputs, as illustrated in Fig. \ref{fig:block_diagram_gen_PreT}.
The AE is then trained with the pre-trained DM until convergence. The main advantage of the pre-training algorithm is the reduction in training time for the E2E framework. Besides, the trained generative model learns $p(\bm{Y}^n|\bm{X}^n)$ for all $\bm{X}^n$ in its domain, enabling its reuse for other applications. A potential challenge is the increased demand for the generative model's learning capacity to comprehensively learn the channel behavior and to generalize effectively to out-of-sample scenarios. Both E2E training algorithms present pros and cons, which we investigate through experiments with different channel models.

Note that the training algorithms apply to any type of generative model and other optimization problems related to the transmitter's part as well. 

\section{Diffusion Models}\label{ssec:diff_model}
This section first explains the DDPM and its modified formulas for conditional generation, as the channel distributions to be learned are conditional on the channel input. Then, two sampling algorithms are explained: one for DDPMs and another for DDIMs with skipped sampling. This is followed by a subsection explaining an improved design of diffusion parameters and parametrization of the DM. Lastly, the NN architecture and an evaluation metric are presented and explained.

\subsection{Diffusion and Denoising Processes}
A DDPM consists of a forward and a reverse diffusion process. Given an input sample from some data distribution $\bm{x}_0\sim q(\bm{x})$, a forward diffusion process can be defined, which progressively adds Gaussian noise to the sample over $T$ time steps, producing increasingly noisy samples $\bm{x}_1,\ldots,\bm{x}_T$ of the original sample. This process is defined as \vspace{-0.1cm}
\begin{align}
    q(\bm{x}_{1:T}|\bm{x}_0)&=\prod_{t=1}^T q(\bm{x}_t|\bm{x}_{t-1})\\
    q(\bm{x}_t|\bm{x}_{t-1})& = \mathcal{N}(\bm{x}_t;\sqrt{1-\beta_t}\bm{x}_{t-1}, \beta_t\mathbf{I}_n),\label{eq:q_diffusion}
\end{align}
where $\{\beta_t\in(0,1)\}_{t=1}^T$. As $T$ approaches infinity, $\bm{x}_T$ converges to an isotropic Gaussian distribution with covariance $\bm{\Sigma}\!=\!\mathbf{I}_n$. From \eqref{eq:q_diffusion}, it should be noted that the resulting noisier sample $\bm{x}_t$ is simply a scaled mean of the previous sample $\bm{x}_{t-1}$ in the process, with additional covariance proportional to $\beta_t$. In other words, for some random vector $\bm{\epsilon}_{t}\!\sim\!\mathcal{N}(\bm{0},\mathbf{I}_n)$ for all $t$, we have
\begin{align}
	\bm{x}_t = \sqrt{1-\beta_t}&\bm{x}_{t-1} +\sqrt{\beta_t} \bm{\epsilon}_{t}.\label{eq:diffusion}
\end{align}
Leveraging the property of the Gaussian distribution that for $\epsilon\sim\mathcal{N}(0,1)$, $\mu+\sigma \epsilon\sim \mathcal{N}(\mu, \sigma^2)$,\footnote{Sometimes referred to as reparametrization trick in the ML literature.} and the sum properties of two Gaussian random variables, one can sample $\bm{x}_t$ directly from $\bm{x}_0$, through recursively applying these properties to arrive at the equation:
\begin{align}
    q(\bm{x}_t|\bm{x}_0)&= \mathcal{N}(\bm{x}_t;\sqrt{\bar{\alpha}_t}\bm{x}_0,(1-\bar{\alpha}_t)\mathbf{I}_n),\label{eq:xt_from_x0} 
\end{align}
where $\alpha_t=1-\beta_t$ and $\bar{\alpha}_t\!=\!\prod_{i=1}^t\alpha_i$, see \cite{ho2020denoising} for details. In another form, we have
\begin{equation}
    \bm{x}_t = \sqrt{\bar{\alpha}_t}\bm{x}_0 + \sqrt{1-\bar{\alpha}_t}\bm{\epsilon}_t,  \label{eq:xt_x0}
\end{equation}
where $\bm{\epsilon}_t \sim \mathcal{N}(\bm{0},\mathbf{I}_n)$. The interesting aspect now is that if we can reverse the process and sample from $q(\bm{x}_{t-1}|\bm{x}_t)$, we can reverse the entire chain and generate data samples from $q(\bm{x})$, by sampling from $\bm{x}_T\!\sim\!\mathcal{N}(\bm{0},\mathbf{I}_n)$ and applying the reverse process. Here, one can use that when $\beta_t \ll 1$ for all $t$, the reverse distribution $q(\bm{x}_{t-1}|\bm{x}_t)$ has the same functional form as the forward one, therefore it is a Gaussian distribution too \cite{sohl2015deep}.
However, this reverse distribution is not given and can be approximated by learning a distribution $p_\theta$ that satisfies the following equations:
\begin{align}
    p_{\theta}(\bm{x}_{0:T})&=p(\bm{x}_T)\prod_{t=1}^T p_{\theta}(\bm{x}_{t-1}|\bm{x}_t),\\
    p_{\theta}(\bm{x}_{t-1}|\bm{x}_t) &= \mathcal{N}(\bm{x}_{t-1};\bm{\mu}_{\theta}(\bm{x}_t,t),\bm{\Sigma}_{\theta}(\bm{x}_t,t)).\label{eq_p_theta}
\end{align}
In simpler terms, the task can be reduced to learning the mean $\bm{\mu}_\theta(\bm{x}_t,t)$ and the covariance $\bm{\Sigma}_\theta(\bm{x}_t,t)$ by optimizing the weights $\theta$ with DL.

\subsection{Training Algorithm}\label{ssec:training_alg_dm}
The training stems from maximizing the likelihood, and often it is replaced by the evidence lower bound (ELBO) as calculating the exact likelihood is prohibitive. The ELBO can be further decomposed into mainly 3 types of terms for $t=1$, $1<t<T$, and $t=T$ as shown in \cite{ho2020denoising}. The commonly-used loss function only considers the term $\mathcal{L}_{t-1}$ for $1<t<T$, which is represented by the Kullbeck-Leibler (KL) divergence between the estimated reverse process $q(\bm{x}_{t-1}|\bm{x}_t,\bm{x}_0)$ and the learned reverse process $p_{\theta}(\bm{x}_{t-1}|\bm{x}_t)$:
\begin{equation}
\mathcal{L}_{t-1}\! =\! \underset{ \bm{x}_t\sim q(\bm{x_t}|\bm{x}_0)}{\mathbb{E}}\!\left[D_{\text{KL}}(q(\bm{x}_{t-1}|\bm{x}_t,\bm{x}_0) || p_{\theta}(\bm{x}_{t-1}|\bm{x}_t))\right].\!\!\! \label{eq_KL_loss}
\end{equation}
Even though the true reverse distribution $q(\bm{x}_{t-1}|\bm{x}_t)$ is unknown, we have $q(\bm{x}_{t-1}|\bm{x}_t,\bm{x}_0)$ when the initial sample $\bm{x}_0$ is given:
\begin{equation}
    q(\bm{x}_{t-1}|\bm{x}_t,\bm{x}_0)= \mathcal{N}(\bm{x}_{t-1};\bm{\mu}_{q}(\bm{x}_t,\bm{x}_0),\bm{\Sigma}_{q}(\bm{x}_t,t))\label{eq:q}
\end{equation}
with
\begin{equation}
    \bm{\mu}_{q}(\bm{x}_t,\bm{x}_0)=\frac{\sqrt{\alpha_t}(1-\bar{\alpha}_{t-1})}{1-\bar{\alpha}_t}\bm{x}_t + \frac{\sqrt{\bar{\alpha}_{t-1}}(1-\alpha_t) }{1-\bar{\alpha}_t}\bm{x}_0\label{eq:x_q}
\end{equation}
and
\begin{equation}
    \bm{\Sigma}_{q}(\bm{x}_t,t)=\frac{(1-\alpha_t)(1-\bar{\alpha}_{t-1})}{1-\bar{\alpha}_t}\mathbf{I}_n=\sigma^2_t\mathbf{I}_n,\label{eq_sigmat}
\end{equation}
where $\sigma_t^2=\frac{\sqrt{1-\bar{\alpha}_{t-1}}}{\sqrt{1-\bar{\alpha}_t}}\beta_t$. Since the covariance is constant, one only needs to learn a parametric mean function to learn the reverse process. It is observed in \cite{ho2020denoising} that $\sigma_t^2=\beta_t$ works as well in experiments, and this can simplify the formula.

If we define $\bm{\Sigma}_\theta(\bm{x}_t,t)$ to match with $\bm{\Sigma}_q(\bm{x}_t,t)$, the optimization of the KL divergence \eqref{eq_KL_loss} of two Gaussian distributions can be simplified as
\begin{equation}
\min_{\theta} \underset{ \bm{x}_0, \bm{x}_t \sim q}{\mathbb{E}}\!\left[ \frac{1}{2\sigma^2_t} ||\bm{\mu}_\theta(\bm{x}_t,t) \!-\! \bm{\mu}_q(\bm{x}_t,\bm{x}_0)||_2^2\right]\label{eq_loss_mu}. 
\end{equation}
From here, we consider $\mathcal{L}_{t-1}$ averaged over $\bm{x}_0\sim q(\bm{x})$ so that it becomes an expectation over the joint distribution of $\bm{x}_0$ and $\bm{x}_t$ and simplify the notation by $\bm{x}_0,\bm{x}_t\sim q$ in the subscript. 
Note that the mean of the learned reverse process $\bm{\mu}_\theta(\bm{x}_t, t)$ does not include the original sample $\bm{x}_0$ as an input, as it is designed to work as a generator during the sampling. Hence, we define $\bm{\mu}_\theta(\bm{x}_t, t)$ in the matched form as $\bm{\mu}_q(\bm{x}_t,\bm{x}_0)$ and use $\hat{\bm{x}}_\theta(\bm{x}_t,t)$ instead of $\bm{x}_0$, which is the predicted value of $\bm{x}_0$ based on the noisy data $\bm{x}_t$ at time step $t$:
\begin{equation}
    \bm{\mu}_{\theta}(\bm{x}_t,t)\coloneqq \frac{\sqrt{\alpha_t}(1-\bar{\alpha}_{t-1})}{1-\bar{\alpha}_t}\bm{x}_t + \frac{\sqrt{\bar{\alpha}_{t-1}}(1-\alpha_t) }{1-\bar{\alpha}_t}\hat{\bm{x}}_{\theta}(\bm{x}_t,t).\label{eq_mu_theta}
\end{equation}
With the matching form, the optimization becomes 
\begin{equation}
\min_{\theta} \underset{ \bm{x}_0, \bm{x}_t \sim q}{\mathbb{E}} \left[\frac{\bar{\alpha}_{t-1}(1-\alpha_t)^2}{2\sigma^2_t(1\!-\!\bar{\alpha}_t)^2}||\hat{\bm{x}}_{\theta}(\bm{x}_t,t) \!-\!\bm{x}_0||_2^2\right].\label{predict_x_0}
\end{equation}

The prediction of the initial sample, $\hat{\bm{x}}_{\theta}$, can be accomplished directly by an NN, referred to as $x$ prediction \cite{nichol2021improved}, or by employing a parametrization technique based on \eqref{eq:xt_x0}. The classical DM \cite{ho2020denoising} uses the latter approach by defining the estimated noise term, $\hat{\bm{\epsilon}}_\theta (\bm{x}_t,t)$, satisfying $\bm{x}_t = \sqrt{\bar{\alpha}_t}\hat{\bm{x}}_{\theta}(\bm{x}_t,t) + \sqrt{1-\bar{\alpha}_t}\hat{\bm{\epsilon}}_{\theta}(\bm{x}_t,t)$ as \eqref{eq:xt_x0}, i.e.,
\begin{align}
\hat{\bm{\epsilon}}_\theta (\bm{x}_t,t) \coloneqq \frac{1}{\sqrt{1-\bar{\alpha}_t}}\left(\bm{x}_t - \sqrt{\bar{\alpha}_t}\hat{\bm{x}}_{\theta}(\bm{x}_t,t)\right).\label{eq_epsilon_theta}
\end{align}
This strategy is called as $\bm{\epsilon}$ prediction, and $\hat{\bm{\epsilon}}_\theta(\bm{x}_t,t)$ indicates the estimated noise in $\bm{x}_t$ as learned by the NN.
Then, the optimization task expressed in \eqref{eq_loss_mu} can be simplified as the following form
\begin{equation}
  \min_{\theta} \underset{\bm{x}_0\sim q, \bm{\epsilon}_t\sim  \mathcal{N}(\bm{0},\mathbf{I}_n)}{\mathbb{E}}\!\! \left[ \frac{(1-\alpha_t)^2}{2\sigma^2_t(1\!-\!\bar{\alpha}_t)\alpha_t}||\hat{\bm{\epsilon}}_{\theta}(\bm{x}_t,t) \!-\!\bm{\epsilon}_t||_2^2\right]. \!
\end{equation} 
We remark that the randomness of $\bm{x}_t$ in the subscript is replaced by $\bm{\epsilon}_t$ because $\bm{x}_t$ is explicitly known for given $\bm{x}_0$ and $\bm{\epsilon}_t$.
Empirical results presented in \cite{ho2020denoising} suggest that this optimization task performs better without the factor in front, leading to the following simple loss function 
\begin{align}
\mathcal{L}_{\bm{\epsilon}} = \underset{t\sim \mathcal{U}\{1,T\}, \bm{x}_0\sim q,  \bm{\epsilon}_t\sim \mathcal{N}(\bm{0},\mathbf{I}_n)}{\mathbb{E}} \left[||\hat{\bm{\epsilon}}_{\theta}(\bm{x}_t,t)-\bm{\epsilon}_t||_2^2\right],\label{eq_loss}
\end{align}
which is averaged over uniformly distributed time $t$. 

\subsection{Conditional Diffusion Models}\label{ssec:conditional_diff}
To use the DM for generating a differentiable channel, two conditions need to be satisfied: 1) the DM should be capable of generating a conditional distribution, and 2) the channel input should serve as a trainable parameter within the DM. 
A channel is characterized by a distribution of the channel output, conditioned on the channel input. For channel generation, the generative model should have the ability to produce multiple conditional distributions $q(\bm{x}_0|\bm{c})$ controlled by a condition $\bm{c}$, which is dependent on the channel input. For simplicity, we assume that the same forward process applies to all potential conditions, i.e., $\beta_t$ does not depend on $\bm{c}$. Conversely, the backward step operates conditionally on $\bm{c}$ as $p_\theta(\bm{x}_{t-1}|\bm{x}_t, \bm{c}) = \mathcal{N}(\bm{x}_{t-1};\bm{\mu}_\theta(\bm{x}_t,t,\bm{c}), \bm{\Sigma}_\theta(\bm{x}_t,t,\bm{c}))$.
With the same choice of the simplified loss as in \eqref{eq_loss}, the loss of the conditional DM becomes
\begin{align}
\mathcal{L}_{\bm{\epsilon},c} =\underset{t, \bm{\epsilon}_t, \bm{c} \sim p(\bm{c}), \bm{x}_0\sim q(\bm{x}_0|\bm{c})}{\mathbb{E}}\left[||\hat{\bm{\epsilon}}_{\theta}(\bm{x}_t,t,\bm{c})\!
-\!\bm{\epsilon}_t||_2^2\right]\label{eq_loss_c},
\end{align}
where $\hat{\bm{\epsilon}}_\theta(\bm{x}_t,t,\bm{c})$ is defined to satisfy 
\begin{align}
\bm{\mu}_\theta(\bm{x}_t,t,\bm{c}) = \frac{1}{\sqrt{\alpha_t}} \bm{x}_t - \frac{1-\alpha_t}{\sqrt{1-\bar{\alpha}_t}\sqrt{\alpha_t}}\hat{\bm{\epsilon}}_\theta(\bm{x}_t,t,\bm{c}) .\label{eq_mu_theta_c}
\end{align}
The random vector $\hat{\bm{\epsilon}}_\theta(\bm{x}_t,t,\bm{c})$ can be learned by using DL.

Regarding channel approximation, our objective is to learn and generate the channel output, denoted as our data point $\bm{x}_0=\bm{Y}^n$, which is conditioned by the channel input $\bm{c}=\bm{X}^n$. 

\subsection{Sampling Algorithms: DDPM and DDIM}\label{ssec:sampling_alg}

Samples are generated by denoising white noise. Starting from $\bm{x}_T$, sampled from $\mathcal{N}(\bm{0},\mathbf{I}_n)$, the denoising process calculates $\bm{x}_0(\bm{c})$ by using the following equation: 
\begin{align}
\bm{x}_{t-1}(\bm{x}_t,t,\bm{c}) &= \bm{\mu}_\theta(\bm{x}_t, t, \bm{c})\! + \!\sigma_t\bm{\epsilon}\\
=&\frac{1}{\sqrt{\alpha_t}} \bm{x}_t \! - \! \frac{1-\alpha_t}{\sqrt{1-\bar{\alpha}_t}\sqrt{\alpha_t}}\hat{\bm{\epsilon}}_\theta(\bm{x}_t,t,\bm{c}) + \sigma_t \bm{\epsilon}_t\label{eq_sampling_DDPM}
\end{align}
for $t=T, T-1, \dots, 1$ with the learned $\hat{\bm{\epsilon}}_\theta(\bm{x}_t,t,\bm{c})$, where $\bm{\epsilon}_t$ is sampled from $\mathcal{N}(\bm{0},\mathbf{I}_n)$ for every $t$. 
This equation is simply obtained by substituting $\bm{\mu}_\theta$ with its definition from \eqref{eq_mu_theta} and replacing $\hat{\bm{x}}_\theta$ with the definition of $\hat{\bm{\epsilon}}_{\theta}$ from \eqref{eq_epsilon_theta}.

A key limitation of DMs stems from their slow sampling speed. To address this problem, we utilize the skipped sampling algorithm and the DDIM, introduced by \cite{song2021denoising}. 
Skipped sampling denoises a sub-sequence $\bm{\tau}$ of $(1,2,\dots,T)$, referred to as the \textit{trajectory}.
Briefly speaking, \cite{song2021denoising} expands the DMs to encapsulate non-Markovian diffusion processes in a parametric form by $\sigma_t$ and the ensuing sample generation. In this generalization, DDPM is a special case that uses a Markovian process when $\sigma_t$ is set to satisfy \eqref{eq_sigmat}. Another notable example is the DDIM, where the diffusion process becomes deterministic given $\bm{x}_0$ and $\bm{x}_{t-1}$, characterized by $\sigma_t=0$, and consequently, its generative process, from $\bm{x}_T$ to $\bm{x}_0$, also becomes deterministic.\footnote{Refer to \cite[Subsection 4.1-2]{song2021denoising} for more details.} DDIMs can be trained by the same loss function as DDPMs, implying that a pre-trained DDPM model could be used for the DDIM sampling algorithm. Compared with DDPMs, DDIMs showed a marginal degradation in sample quality for skipped sampling in \cite{song2021denoising}. As such, we consider only DDIMs for skipped sampling in this paper.
Let $S\coloneqq\dim(\bm{\tau})$ and $\bm{\tau}\coloneqq (\tau_1, \tau_2, \dots, \tau_S)$. 
In a parametric form, the denoising step can be written as
\begin{align}
&\bm{x}_{\tau_{i-1}}(\bm{x}_{\tau_i},\tau_i,\bm{c}) \\
&\quad=\sqrt{\bar{\alpha}_{\tau_{i\!-\!1}}}\hat{\bm{x}}_\theta(\bm{x}_{\tau_i},\tau_i,\bm{c}) + \sqrt{1\!-\!\bar{\alpha}_{\tau_{i\!-\!1}}} \hat{\bm{\epsilon}}_\theta(\bm{x}_{\tau_i},\tau_i,\bm{c}), \\
&\hat{\bm{x}}_\theta(\bm{x}_{\tau_i},\tau_i,\bm{c}) =\! \frac{1}{\sqrt{\bar{\alpha}_{\tau_i}}}\left(\bm{x}_{\tau_i} \!-\! \sqrt{1-\bar{\alpha}_{\tau_i}} \hat{\bm{\epsilon}}_\theta(\bm{x}_{\tau_i},\tau_i,\bm{c})\right)\label{eq_epsilon_DDIM}
\end{align}
for $i=1,2 \dots, S$, where $\tau_0=0$ and $\tau_S = T$. The first equation is from \cite{song2021denoising}, and the second equation is from \eqref{eq_epsilon_theta}.
This can be solved as the following recursive formula:
\begin{align}
&\bm{x}_{\tau_{i-1}}(\bm{x}_{\tau_i},\tau_i,\bm{c}) \\
=& \frac{\sqrt{\bar{\alpha}_{\tau_{i\!-\!1}}}}{\sqrt{\bar{\alpha}_{\tau_i}}}\bm{x}_{\tau_i} \!-\! \left( 
 \frac{\sqrt{\bar{\alpha}_{\tau_{i\!-\!1}}} \sqrt{1\!-\!\bar{\alpha}_{\tau_i}}}{\sqrt{\bar{\alpha}_{\tau_i}}} \!-\! \sqrt{1\!-\!\bar{\alpha}_{\tau_{i\!-\!1}}}   \right)\hat{\bm{\epsilon}}_\theta(\bm{x}_{\tau_i},\tau_i,\bm{c}).\label{eq_epsilon_DDIM_rec}
\end{align}
 Note that the equation does not have a random term compared to \eqref{eq_sampling_DDPM}, and the coefficients are not simply canceled out like \eqref{eq_sampling_DDPM} because of the skipped denoising steps. 

\begin{figure}
    \centering
    \includegraphics[width=0.95\columnwidth]{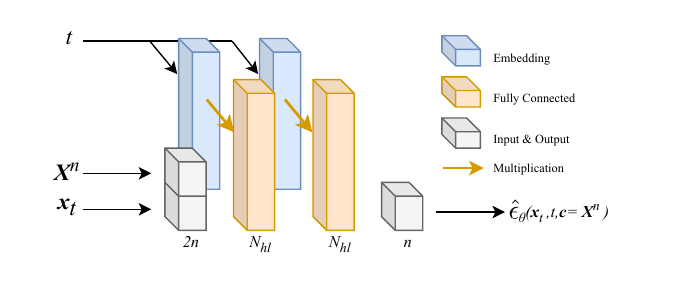}
    \caption{The architecture of the DM is visualized.}
    \label{fig:diff_arch}
\end{figure}

\subsection{Diffusion Noise Scheduling for Zero SNR and $v$ Prediction}\label{ssec:0-SNR}
A recent work \cite{lin2023common} pointed out that for finite $T$ and small $\beta_t$, the diffused sample distribution has a nontrivial difference from the white noise distribution. This causes the signal-to-noise ratio (SNR) of the diffusion to be nonzero. More specifically, from \eqref{eq:xt_x0}, the SNR of diffusion is given by $\frac{\bar{\alpha}_T}{1-\bar{\alpha}_T}$; for instance, for $T=100$ and $\beta_t=0.05$ for all $t$, the SNR equals $5.96\times 10^{-3}$. Although this is a relatively small value, it is nonetheless nonzero. 
In the field of computer vision, eliminating this discrepancy could improve the coverage of samples, as suggested by \cite{lin2023common}. One proposed solution is to enforce a 0-SNR diffusion by setting $(1-\bar{\alpha}_T)\!=\!1$ in \eqref{eq:xt_from_x0}, implying $\alpha_T\!=\!\bar{\alpha}_T\!=\!0$ and $\beta_T\!=\!1$.
However, this adjustment needs a change in parametrization because the diffusion step \eqref{eq:diffusion} at $t\!=\!T$ is simplified to $\bm{x}_T\sim\mathcal{N}(\bm{0},\mathbf{I}_n)$, having no information about $\bm{x}_{T-1}$ nor $\bm{x}_{0}$. 
More technically, $\bm{x}_{T-1}$ is obtained upon the prediction of the mean of the denoising distribution as in \eqref{eq_mu_theta} through $\hat{\bm{x}}_\theta(\bm{x}_T, T)$, but $\hat{\bm{x}}_\theta(\bm{x}_T, T)$ is unobtainable for given $\hat{\bm{\epsilon}}_\theta (\bm{x}_T,T)$ because it is canceled out due to the zero coefficient in \eqref{eq_epsilon_theta}. 
Hence, to learn the distribution $q(\bm{x}_{T-1}|\bm{x}_0)$, we require a new parametrization that includes direct prediction of $\bm{x}_0$ by using the NN, such as $x$ prediction \cite{nichol2021improved} or $v$ prediction \cite{salimans2022progressive}. In this paper, we use $v$ prediction to enable 0-SNR $\beta_t$ scheduling. The vector $v$ is a combination of $\bm{\epsilon}$ and $\bm{x}_0$, defined as $v(\bm{\epsilon}, \bm{x}_0)\coloneqq \sqrt{\bar{\alpha}_t} \bm{\epsilon} - \sqrt{1-\bar{\alpha}_t}\bm{x}_0$, so its prediction $\hat{\bm{v}}_\theta(\bm{x}_t, t)$ is defined as
\begin{equation}
\hat{\bm{v}}_\theta(\bm{x}_t, t)\coloneqq \sqrt{\bar{\alpha}_t} \hat{\bm{\epsilon}}_{\theta}(\bm{x}_t,t) - \sqrt{1-\bar{\alpha}_t} \hat{\bm{x}}_\theta(\bm{x}_t,t).
\end{equation}
An NN is then trained to learn $\hat{\bm{v}}_\theta$ by using the following loss function:
\begin{equation}
\mathcal{L}_{\bm{v}} = \mathbb{E}_{t, \bm{\epsilon}, \bm{x}_0}\left[||\hat{\bm{v}}_\theta(\bm{x}_t,t)-\bm{v}(\bm{\epsilon}, \bm{x}_0)||_2^2\right]. 
\end{equation}
For sample generation in the DDPM framework, one can use the following denoising equation:
\begin{align}
\bm{x}_{t-1} =  \sqrt{\alpha_t}\bm{x}_t - \frac{\sqrt{\bar{\alpha}_{t-1}}(1-\alpha_t) }{\sqrt{1-\bar{\alpha}_t}} \hat{\bm{v}}_\theta(\bm{x}_t,t)+ \sigma_t \bm{\epsilon}.\label{eq_ddpm_v}
\end{align}
In the context of DDIM, the denoising equation is:
\begin{align}
\bm{x}_{\tau_{i-1}} &=\left(\sqrt{\bar{\alpha}_{\tau_{i-1}}\bar{\alpha}_{\tau_{i}}} + \sqrt{(1-\bar{\alpha}_{\tau_{i-1}})(1-\bar{\alpha}_{\tau_{i}})} \right)\bm{x}_{\tau_i} \\
&\ +\left( \! \sqrt{\bar{\alpha}_{\tau_{i}}(1 \!- \!\bar{\alpha}_{\tau_{i-1}})} \!-\! \sqrt{\bar{\alpha}_{\tau_{i-1}}(1-\bar{\alpha}_{\tau_{i}})} \right)\hat{\bm{v}}_\theta(\bm{x}_{\tau_i},\tau_i).
\end{align}
The proofs of these denoising equations can be found in Appendix \ref{appdx_sampling_v_prediction}. The inclusion of the condition $\bm{c}$ can be accomplished similarly to the process detailed in Section~\ref{ssec:conditional_diff}.

\subsection{Neural Network Architecture}\label{ssec:NN_architecture}
For generating images like CIFAR10, CelebA, and LSUN, U-Nets \cite{ronneberger2015u} are typically employed in DMs \cite{ho2020denoising}. A U-Net first down-samples the input and then up-samples it, leveraging some skip connections. This forms a U-shaped block diagram, using convolutional layers for both down-sampling and up-sampling. U-Net could be beneficial for a high-dimensional channel where there is interference between dimensions. However, in this paper, we mostly use simple memoryless channel models with a small block length $n$. We utilize basic linear NNs for those channel models and show the potential of 1D U-Net for a correlated distribution briefly. 

Fig. \ref{fig:diff_arch} illustrates the generic NN structure used in this paper. It is a feed-forward NN of two fully-connected hidden layers, each having $N_{hl}$ nodes. The input layer takes a noisy sample $\bm{x}_t$ and the channel input $\bm{X}^n$, and the NN produces the noise component $\hat{\bm{\epsilon}}_\theta(\bm{x}_t,t,\bm{c}=\bm{X}^n)$. 
Instead of training $T$ distinct models for each time step, the NN typically includes embedding layers for $t$ to function across all $t$. We define an embedding layer of $t$ for each hidden layer with the matching size and multiply it to the hidden layer as demonstrated in Fig. \ref{fig:diff_arch}. The result of the multiplication is activated by the Softplus function for every hidden layer. The output layer is a linear layer. 

We feed the channel input $x$ into the NN as an additional input for two main reasons: to conditionally generate the channel output via the DM and to ensure that the generated channel output is differentiable with respect to the channel input. This last aspect allows us to form a differentiable backpropagation path for the encoder NN. More specifically, in the DM, the backward step $p_\theta(\bm{x}_{t-1}|\bm{x}_t,\bm{c})$ follows a normal distribution. The mean of this distribution satisfies \eqref{eq_mu_theta_c}, which only involves scaling and addition of $\bm{x}_t$ and $\hat{\bm{\epsilon}}_\theta(\bm{x}_t,t,\bm{c})$. With this construction, the chain rule can be applied without issues when obtaining the partial derivatives of the channel output with respect to the channel input, i.e., during backpropagation for training the encoder NN.

\subsection{Sliced Wasserstein Distance as a Performance Metric}
To measure the quality of the generated distributions, we use the sliced Wasserstein distance (SWD) between the ground truth and the generated distributions, as suggested by \cite{rabin2012wasserstein}. 
Briefly, the $p$-Wasserstein distance $W_p(X,Y)$ measures the distance between two probability distributions $P(X)$ and $Q(Y)$ by using $N$ samples of them as
\begin{equation}
    W_p(X,Y)=\inf_{\pi}\left(\frac{1}{N}\sum_{i=1}^N||X_i - Y_{\pi(i)} ||^p\right)^{\frac{1}{p}},
\end{equation}
where $\pi$ is an arbitrary permutation of $N$ elements. Calculating the exact value for large $N$ and multidimensional samples is often prohibitive due to the infimum over all possible permutations. The optimal permutation is analytically known only for one-dimensional (1D) random variables. Thus, for multidimensional variables, the Wasserstein distance is often approximated to avoid computation heavy simulations. The SWD is one of them, approximating the Wasserstein distance by using linear projections, called the Radon transform, onto one-dimensional space. For $p=1$, the SWD is defined as
\begin{equation}
    \tilde{W}_1(X,Y)=\int_{\theta\in\Omega} \tilde{W}_1(X_\theta,Y_\theta) \text{ d}\theta,  
\end{equation}
where $X_{\theta}=\{\langle X_i,\theta \rangle \}_{i=1}^{N} \in\mathbb{R}$ and $\Omega=\{\theta\in\mathbb{R}^d;||\theta||^2=1\}$ is the unit sphere. By using the 1D optimal permutation, this can be re-written as 
\begin{equation}
    \tilde{W}_1(X,Y)=\int_{\theta\in\Omega} \frac{1}{N}\sum_{i=1}^N |\langle X_i - Y_{\pi^*_\theta(i)}, \theta\rangle | \text{ d}\theta,  
\end{equation}
where $\pi^*_\theta$ is the optimal permutation for given projection $\theta$ that matches the $j$-th largest value in $X_\theta$ to the $j$-th largest value in $Y_\theta$ for all $j=1,2,\dots,N$.

\begin{figure*}[h!]
    \centering
    \begin{subfigure}{0.9\columnwidth}
        \centering
        \includegraphics[width=\columnwidth]{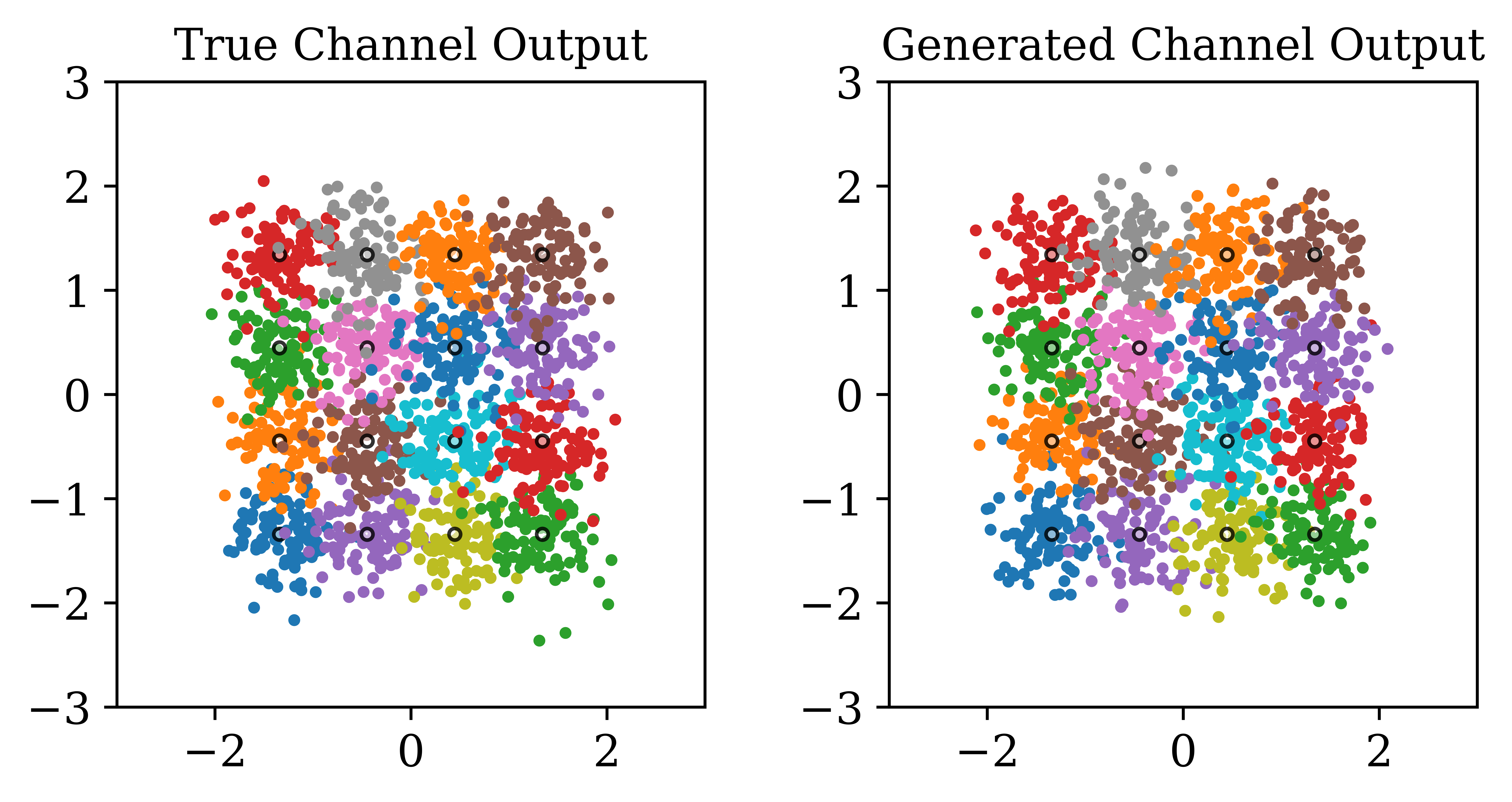}
    \end{subfigure}
    \begin{subfigure}{1.1\columnwidth}
        \centering
        \includegraphics[width=\columnwidth]{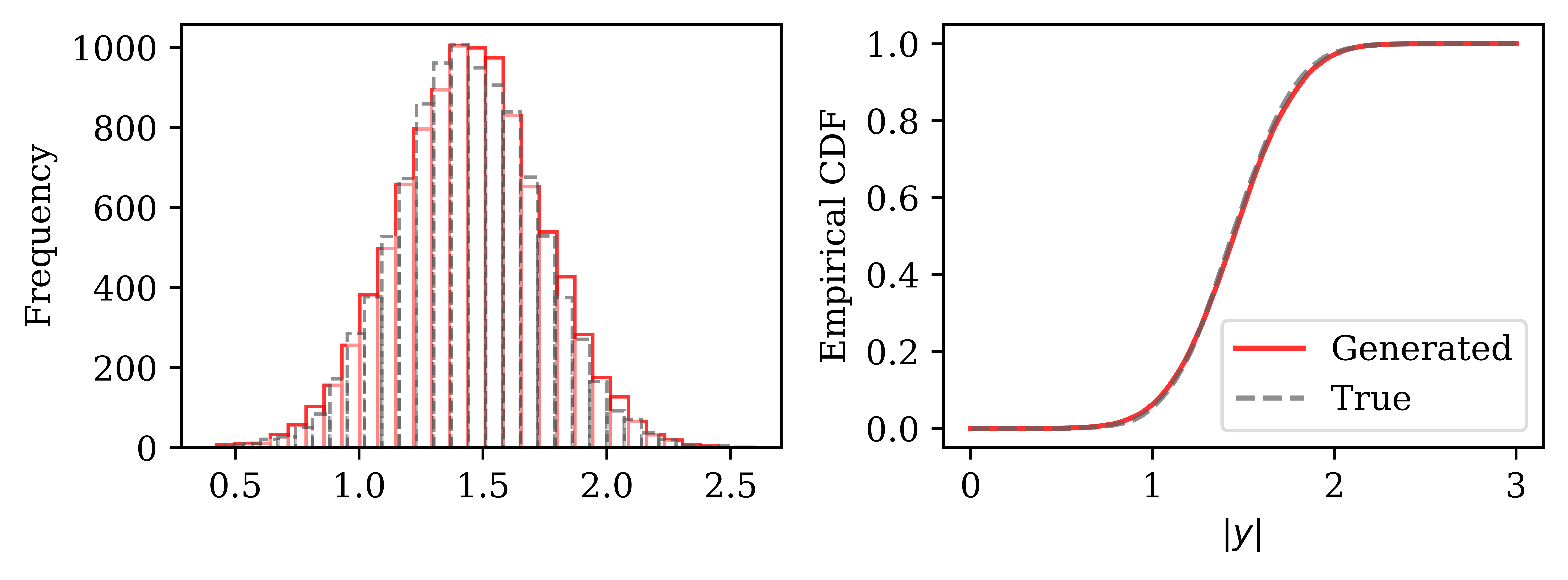}
    \end{subfigure}
    \caption{The generative performance of the DM is tested for an AWGN channel with $16$-QAM modulated symbols and $E_b/N_0=\qty{5}{\decibel}$. The constellation of channel output is drawn with $100$ samples for each message. For $m=4$, the histogram and empirical CDF of the channel output's norm are obtained with $10^4$ samples and compared. In the histogram, the frequency indicates the occurrence of the values in the interval among $10^4$ samples.  }\label{fig:Constell_AWGN_QAM16}
\end{figure*}

\section{Simulation Results}
We evaluate first the channel generation of DMs and then the E2E communication performance with the DMs across different channel scenarios: an AWGN channel, a real Rayleigh fading channel, and a solid state power amplification (SSPA) channel. In all experiments, we assume the messages to be uniformly distributed. The architectures of the AEs and DMs follow the structure detailed in Section~\ref{ssec:AE_cDiff}, but we individually tune the number of hidden layers and hidden nodes for each channel model. For hyperparameters, $T=100$ is used for all DMs, and we use the optimal $\beta_t$ scheduling identified through our hyperparameter search in Section~\ref{ssec:beta_scheduling}. 

\subsection{Channel Generation: 16-QAM for AWGN Channel}\label{ssec:ch_gen_test}
This subsection demonstrates the channel generative performance of the conditional DM through an experiment with a 16-ary quadrature amplitude modulation (QAM) and an AWGN channel with $E_b/N_0=\qty{5}{\decibel}$.
The AWGN channel is defined as $\bm{Y}^n = \bm{X}^n + \bm{Z}^n$, where $\bm{X}^n\in\mathbb{R}^n$ is the channel input, and $\bm{Z}^n$ is the independent channel noise that follows $\bm{Z}^n\sim \mathcal{N}(0,\sigma^2\mathbf{I}_n)$ for some $\sigma>0$. We employed normalized 16-QAM symbols for the channel input and set the $\sigma$ value such that $E_b/N_0=\qty{5}{\decibel}$.

For the diffusion process, we used a sigmoid-scheduled $\beta_t$ increasing from $10^{-3}$ to $0.05$, i.e., $\beta_t= 0.001 + 0.05\left(\frac{1}{1+\exp(-(1+12(t-1)/T))}\right)$. We implemented the conditional DM outlined in Section~\ref{ssec:NN_architecture}, with $N_{hl}=110$. The DM was optimized by the accelerated adaptive moment estimation (Adam) optimizer \cite{kingma2014adam}, with a learning rate of $10^{-3}$, dataset size of $10^6$, batch size of $100$, and 5 epochs. We used channel inputs following the standard isotropic normal distribution and their channel outputs as the training dataset, consistent with the pre-training algorithm.

Fig.~\ref{fig:Constell_AWGN_QAM16} compares the constellations of the channel outputs of the channel model and the generated channel outputs (in the first two subfigures from the left), demonstrating notably similar distributions. The conditioning capability is proved by the channel output distributions generated for each message. To further verify the distribution, we generated the histogram and empirical cumulative density function (CDF) of the channel output's magnitude, using $10^4$ samples for each message. To conserve space, only the graphs corresponding to $m=4$ are shown in Fig.~\ref{fig:Constell_AWGN_QAM16}. Both graphs underscore that the generated distribution mirrors the true distribution precisely, with analogous results found for other messages as well.

In \cite{Kim2023diff}, a similar experiment is conducted with a different DM and iterative training algorithm. The NN of \cite{Kim2023diff} has embedding layers of $m$ multiplied to the hidden layers, and it was trained with the set of encoded symbols and their respective channel outputs. A comparison to \cite{Kim2023diff} proves that it is possible to generate high-quality samples by training with normally distributed channel inputs and an NN without the embedding layers of $m$. This effectively demonstrates the potential of the pre-training algorithm to serve as a substitute for the encoder-specific iterative training algorithm.
 
\begin{figure*}
    \centering
    \begin{subfigure}{0.63\columnwidth}
    \centering
    \includegraphics[width=\textwidth]{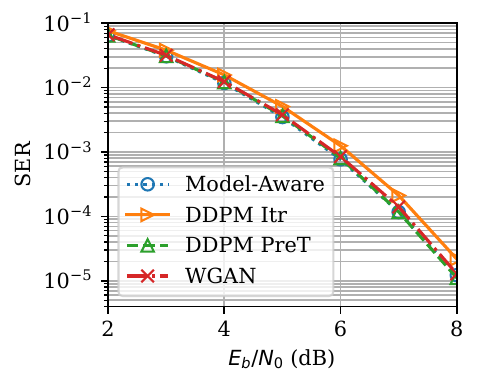}
    \subcaption{An AWGN channel.}
    \label{fig:SER_AWGN}
    \end{subfigure}
    \centering
    \begin{subfigure}{0.7\columnwidth}
    \centering
    \includegraphics[width=\textwidth]{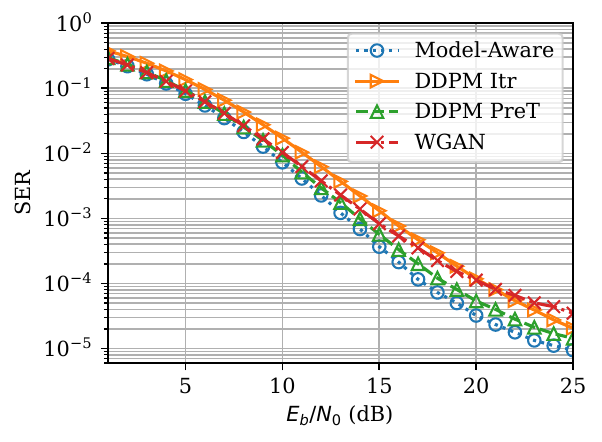}
    \subcaption{A Rayleigh fading channel.}
    \label{fig:SER_Rayleigh}
    \end{subfigure}
    \begin{subfigure}{0.63\columnwidth}
    \centering
    \includegraphics[width=\textwidth]{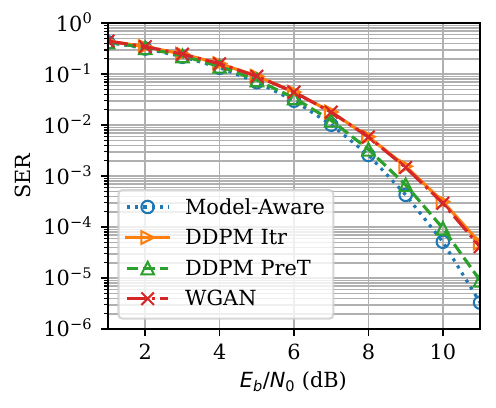}
    \caption{An SSPA channel.}
    \label{fig:SER_SSPA}
    \end{subfigure}
    \centering
    \caption{The E2E SER for varying $E_b/N_0$ values is tested for the 3 channel models.}\label{fig:SER}
\end{figure*}

\subsection{E2E Learning: AWGN Channel}\label{ssec:AWGN_channel}

\begin{table}
\caption{The Architecture of the AE used for Simulations.}
\label{Tbl:AE_simple}
\begin{center}
\begin{tabular}{ c | c| c| c| c| c| c}

\hline
 & \multicolumn{3}{c|}{Encoder} & \multicolumn{3}{c}{Decoder}  \\
 \hline
layer & 1 & 2 & 3 & 1 & 2-3 & 4 \\
 neurons & M & M & n &n & M & M \\
activation & - & ELU & - & - & ELU & - \\
\hline
\end{tabular} 

\end{center} 
\end{table}

We evaluate the SERs of the E2E frameworks, where the AE is trained with either an AWGN channel model or learned channels by generative models.
For this simulation, we set $n=7$, $M=16$ and trained the models at $E_b/N_0=\qty{5}{\decibel}$. 
As AWGN channels are simple with only additive Gaussian noise, we employed straightforward NN models to define the encoder and the decoder. These models consist of linear layers with $M=16$ neurons, activated by exponential linear unit (ELU) activation. Table~\ref{Tbl:AE_simple} provides further details on the architecture. The encoder and the decoder were trained simultaneously by using a Nesterov-accelerated Adam (NAdam) optimizer with learning rate equal to $10^{-3}$.

 Incorporating the outcome of the preceding study \cite{Kim2023diff}, we utilize the result of a DDPM trained through an iterative training algorithm from \cite{Kim2023diff}. For pre-training, we apply the DM architecture defined in Section~\ref{ssec:NN_architecture}, with $N_{hl}=110$ hidden neurons for each layer, and cosine scheduled $\beta_t$, as proposed in \cite{nichol2021improved}. This setup prescribes that $\bar{\alpha}_t=f(t)/f(0)$ where $f(t)=\cos\left(\frac{t/T}{1}\cdot\frac{\pi}{2}\right)^2$. Note that $\bar{\alpha}_T=\cos(\pi/2)^2=0$, and the $v$ prediction, as explained in Section~\ref{ssec:0-SNR}, should be used. 
 The training uses an Adam optimizer with a learning rate of $10^{-4}$, a dataset size of $10^7$, a large batch size of $5000$ to minimize intra-batch variance from the true distribution, and $30$ epochs. The AE is trained for $50$ epochs using a dataset of $10^6$ samples and a batch size $100$. 
  
The E2E SER is evaluated for integer values of $E_b/N_0$ from $\qtyrange{2}{8}{\decibel}$ and is illustrated in Fig.~\ref{fig:SER_AWGN}. The legends `DDPM Itr' and `DDPM PreT' denote the iterative and pre-training algorithms, respectively. In addition, we implemented two other E2E frameworks: Model-Aware and WGAN.
Model-Aware directly uses the simulated channel output for training the AE, and its SER curve represents the target to achieve with the generated channels. In \cite{o2017introduction}, it was shown that the Model-Aware case can achieve the same SER as a Hamming code with the D/BPSK and maximum likelihood decoding, hence we consider the Model-Aware case as the baseline in our study.  
For the WGAN's generator and critic, we employed NNs with two linear hidden layers, each with 128 nodes, and activated by the rectified linear unit (ReLU). Both the generator and the critic models' input layers receive the channel input, ensuring the differentiability of the generated channel. The models are trained by the pre-training algorithm.
We used a root-mean-square propagation (RMSprop) optimizer for them with learning rate $10^{-4}$ and batch size $5000$ across $2\times10^4$ iterations. The AE is then trained for 10 epochs using a dataset size of $10^7$. 

Fig.~\ref{fig:SER_AWGN} shows that SER values decrease with increasing $E_b/N_0$, yielding concave curves. Both DDPM PreT and the WGAN exhibit SER values close to the target curve, whereas DDPM Itr still has a thin gap to the optimal curve.

\subsection{E2E Learning: Rayleigh Fading Channel}
 We consider a real Rayleigh fading channel, defined by $\bm{Y}^n = \bm{H}^n\circ \bm{X}^n + \bm{Z}^n$, where $\circ$ symbolizes the element-wise multiplication operator, and $\bm{H}^n$ satisfies the condition $\Pr(H_i=x;\sigma_R)=\frac{x}{\sigma_R^2}\exp(-x^2/(2\sigma_R^2))$ for $x>0$, $\sigma_R>0$ for all $i=1,2,\dots,n$, and $\bm{Z}^n\sim\mathcal{N}(0,\sigma^2\mathbf{I})$ for some $\sigma>0$. We assume a real Rayleigh fading channel with $n=7$, $\sigma_R=1$ for $M=16$ messages, and the models 
 were trained at $E_b/N_0=\qty{12}{\decibel}$. The AE model's architecture is consistent with the one used for the AWGN channel in Section~\ref{ssec:AWGN_channel}.

 Fig.~\ref{fig:SER_Rayleigh} presents the test SER values of the E2E frameworks similarly in Section~\ref{ssec:AWGN_channel}. 
 The DM for the pre-training algorithm, defined with $N_{hl}=128$, cosine-scheduled $\beta_t$ was trained using the $v$ prediction method. The training parameters involved a dataset size of $10^7$, batch size of $5000$, $30$ epochs, and a learning rate of $10^{-4}$. The AE was trained with a dataset size of $10^6$, batch size of $100$, an initial learning rate of $10^{-3}$ for the first $30$ epochs, and then reduced to $10^{-4}$ for the subsequent $60$ epochs. For each epoch, the decoder and encoder were trained successively. 

 The WGAN model was similarly defined to the one described in Section~\ref{ssec:AWGN_channel} and differs only in having 256 neurons per hidden layer and a learning rate set at $\num{5e-5}$. For the WGAN and the Model-Aware cases, the AE was trained with a dataset of $10^7$ samples, a batch size of $1000$, a learning rate of $10^{-3}$, and $10$ epochs. 

 Model testing was carried out for integer $E_b/N_0$ values ranging from \qtyrange{1}{25}{\decibel}. The DDPM PreT case outperforms DDPM Itr and WGAN cases, achieving an almost optimal curve.  
 In the semi-log scale plot, the SER curves of DDPMs and Model-Aware frameworks decrease linearly, while the WGAN curve, maintaining a minuscule gap, aligns with the optimal curve until around \qty{17}{\decibel}, where it begins to diverge. Such a limitation is consistent with the GAN results observed in \cite{ProductAE}. 

\subsection{E2E Learning: SSPA Channel}
In order to test the performance of DMs on a nonlinear channel, we utilized the SSPA channel model proposed by \cite{rapp1991effects}. In this model, an $n_c$-dimensional complex channel input signal $\bm{X}^{n_c}\in\mathbb{C}^{n_c}$, characterizes the SSPA channel as $\bm{Y}^{n_c}=P^{n_c}\left(\bm{X}^{n_c}\right)\circ \bm{X}^{n_c}+\bm{Z}^{n_c}$, where $P^{n_c}\left(\bm{X}^{n_c}\right)\!=\!(P(|X_1|), P(|X_2|), \dots, P(|X_{n_c}|)) \!\in \!\mathbb{R}^{n_c}$ is a nonlinear power amplification, and each element $Z_i$ of $\bm{Z}^{n_c}$ satisfies $Z_i\sim \mathcal{CN}(0, \sigma^2)$.
The power amplification function of an SSPA model is defined as 
\begin{equation}
    P\left(|X_i|\right) ={v_0}/{\left( 1+ \left[ \frac{v_0\lvert X_i\rvert}{A_0} \right]^{2p}\right)^{\frac{1}{2p}}}
\end{equation} for some $p>0$, $A_0\geq 0$, and $v_0\geq 0$ for all $i=1,2,\dots,n_c$. This function amplifies the signal proportionally to $v_0$, and the amplification becomes saturated when $\lvert X_i\rvert$ is approximately equal to $A_0$. The saturation level's smoothness is determined by $p$: a smaller $p$ results in smoother saturation. For $p\to\infty$, the curve of $P(X_i)$ increases linearly up to $\lvert X_i \rvert = A_0$ and stays constant for $\lvert X_i \rvert > A_0$. This saturation behavior induces the amplification nonlinearity as $P(|X_i|+a|X_j|)\neq P(|X_i|)+aP(|X_j|)$ for arbitrary $|X_i|$, $|X_j|$, and $a\in\mathbb{R}$. 

For simulations, parameters were set to be $M=64$, $n_c=4$, $p=3$, $A_0=1.5$, $v_0=5$, and $E_b/N_0=\qty{8}{\decibel}$ for training. We used $n=2n_c$ real values to represent $n_c$-dimensional complex vectors. Fig.~\ref{fig:SER_SSPA} illustrates the test SER curves of the four frameworks. The AEs in all scenarios follow the specifications outlined in Table~\ref{Tbl:AE_simple}.

For the Model-Aware case, training was done with a dataset size of $10^6$, a batch size of $1024$, and learning rates decreasing down to $10^{-5}$ for 160 epochs in total. The DMs architecture mirrored the one used for an AWGN channel in Section~\ref{ssec:AWGN_channel}. 
From a hyperparameter search, the 0-SNR cosine-scheduled $\beta_t$ and the corresponding $v$ prediction for denoising were found to be the most effective. The DDPM PreT case is trained with a dataset size of $10^7$, a batch size of $4096$, and learning rate decreasing down to $10^{-6}$ for $160$ epochs in total. Thereafter, the AE is trained with dataset size $10^6$, batch size 128, learning rate $10^{-4}$ for $60$ epochs. The iterative algorithm is evaluated for the same DM as the pretraining algorithm, for simplicity.

The same WGAN model with 256 hidden nodes was employed as for the Rayleigh fading channel, with modifications made to the input and output layer sizes. A WGAN with 128 hidden nodes was also tested, but a noticeable performance gap was found compared to the larger model. The WGAN was trained with a batch size of $5000$ and learning rates decreasing down to $10^{-6}$ for $1.5\times10^5$ iterations in total. Thereafter, the AE was trained with a dataset size of $10^8$, a batch size of $1000$, and the learning rate decreasing down to $10^{-6}$ for $160$ epochs in total.  

Fig.~\ref{fig:SER_SSPA} indicates that the DDPM PreT achieves a near-optimal performance, while the WGAN and DDPM Itr still have clear gaps to the target curve. This suggests that a DM can effectively learn a nonlinear channel distribution as a black box and enable the neural encoder in the E2E framework to learn a near-optimal codebook. 

\begin{table}[!t]
\renewcommand{\arraystretch}{1.3}
\caption{Dimensionality of the NNs in the Experiments.}
\label{Tbl:dimensionality}
\centering
\begin{tabular}{ c | c| c| c}
\hline
    & AWGN & Rayleigh & SSPA \\
 \hline
DDPM Itr & 22\ 407  & 22\ 407 & 60\ 178  \\
\hline
DDPM PreT & 36\ 637 & 74\ 247 & 60\ 178  \\
\hline
WGAN G & 17\ 410 & 67\ 586 & 72\ 200 \\
WGAN C & 17\ 410 & 67\ 329 & 70\ 401 \\
\hline
\end{tabular} 
\end{table}

\subsection{Comparison of the Two Training Algorithms}
For all channel models in Fig.~\ref{fig:SER}, the DDPM with the pre-training algorithm proves its potential to support the E2E learning framework, outperforming the DDPM with the iterative training algorithm and achieving near-optimal performances. 
This suggests the DM's ability to learn not only $M$ distinct distributions but also the general channel behaviour for an arbitrary input. 
This would be especially important for the nonlinear channels because interpolation or extrapolation by the finite $M$ distributions would not be as easy due to the nonlinearity in case the iterative training algorithm is used. The encoder-specific generative model does not necessarily guarantee this performance as it is only trained with finite conditional distributions.\\

In addition to the generative performance, the pre-training algorithm is preferred to the iterative algorithm, as the iterative algorithm often takes longer time and exhibits variance in the reproduction of results. These drawbacks mainly stem from the iterative optimization of a generative model and an autoencoder. 
While the results are significant, it is important to caution that these do not guarantee superior performance from the pre-training algorithm in every instance. The optimal design and hyperparameters of the DDPM differ across the two algorithms, complicating direct comparisons. However, the findings remain meaningful because a comparable or even better performance can be achieved in a relatively effortless manner with the pre-training algorithm.
Thus, for the rest of the simulations, DDPM PreT is used as the representative of DMs. 

\begin{table*}[t!]
\renewcommand{\arraystretch}{1.3}
\caption{SWDs measured for the 3 channel models and the 3 noise scheduling methods.}
\label{Tbl:SWD}
\centering
\begin{tabular}{c|c|c|c|c|c|c|c|c}
\hline
     \multicolumn{9}{c}{AWGN}  \\
 \hline
Scheduling & DDPM & DDIM-100 & DDIM-50 & DDIM-20 & DDIM-10 & DDIM-5 & DDIM-2  & WGAN \\
\hline
constant & 0.016 & 0.083 & 0.105 & 0.136 & 0.165 & 0.152 & 0.151 & \multirow{3}{*}{0.013} \\
\cline{1-8}
sigmoid &0.033&0.152&0.154&0.163  &  0.171&0.174&0.161 &\\
\cline{1-8}
\textbf{cosine} & 0.012&\textbf{0.008}&0.011&0.022&0.043&0.074&   0.135 &\\
\hline
 \end{tabular} 
 \\
 \begin{tabular}{c|c|c|c|c|c|c|c|c}
\hline
     \multicolumn{9}{c}{Rayleigh}  \\
 \hline
Scheduling & DDPM & DDIM-100 & DDIM-50 & DDIM-20 & DDIM-10 & DDIM-5 & DDIM-2 & WGAN \\
\hline
constant & 0.085&0.061&0.063&0.073&0.085&0.109&0.165 & \multirow{3}{*}{0.019} \\
\cline{1-8}
sigmoid &0.046&0.188&0.195&0.208&0.247&0.306&0.475& \\
\cline{1-8}
\textbf{cosine} & 0.013&\textbf{0.008}&0.011&0.024&0.043&0.073&0.123 &\\
\hline
 \end{tabular} 
  \\
 \begin{tabular}{c|c|c|c|c|c|c|c|c}
\hline
     \multicolumn{9}{c}{SSPA}  \\
 \hline
Scheduling & DDPM & DDIM-100 & DDIM-50 & DDIM-20 & DDIM-10 & DDIM-5 & DDIM-2 & WGAN \\
\hline
constant &  0.074&0.132&0.152&0.191&0.230&0.264&0.284& \multirow{3}{*}{0.104} \\
\cline{1-8}
sigmoid &  0.085&0.202&0.208&0.218&0.233&0.261&0.295&  \\
\cline{1-8}
\textbf{cosine} &  0.009&\textbf{0.007}&0.009&0.018&0.032&0.055&0.088&  \\
\hline

 \end{tabular} 
\end{table*}
\subsection{Model Complexity Analysis}
Table~\ref{Tbl:dimensionality} presents the number of parameters utilized in each simulation. By rough ablation study during training, we observed that to learn the Rayleigh fading channel and the SSPA channel for E2E learning, both the DDPM and the WGAN required larger NNs compared to the AWGN channel. This adjustment aligns with expectations, as the AWGN channel output resembles the random seed of generative models, which is generally easier to generate than the other two channel models.

Compared to the iterative training algorithm of DDPM, the pre-training algorithm required larger networks to achieve comparable SERs for both the AWGN channel and the Rayleigh fading channel. For iterative training, the generative model only needs to learn $M$ distributions, whereas, for pre-training, it must learn a general channel distribution for any given channel input $\bm{X}^n$. As such, this increased model complexity seems unavoidable yet justifiable when seeking more stable training and lower SERs. For the SSPA channel, we optimized a DDPM for the pre-training algorithm and used the same NN for the iterative training algorithm. Yet, the SER performance of the DDPM Itr in Fig.~\ref{fig:SER_SSPA} is observed not as good as the DDPM PreT, which shows that the performance gap is not rooted from the model size. 

In this paper, the WGANs are assessed solely with the pre-training algorithm. The DDPM with the pre-training algorithm required a similar or less number of parameters compared to the WGAN, while achieving better E2E SER. This can be explained by the two models' different structure. The DM breaks down a complicated estimation problem into many simple problems by using the gradual denoising, and the NN of a DM learns only one denoising step at once. On the other hand, the generator of a GAN should do the same job at one forward step of the NN, which possibly necessitates a larger and more sophisticated NN.

\section{Performance Enhancement}
To address the slow sampling issue inherent in DMs, we investigate the potential time savings and the consequent impact on sample quality when using DDIMs. This subsection analyzes the trade-off between sampling time and sample quality. We accomplish this by comparing the SWDs and the E2E communication quality achieved with DDPMs and DDIMs. Furthermore, we explore optimization of the $\beta_t$ parameter as a strategy to improve this trade-off. 

\begin{figure*}[h!]
    \centering
    \begin{subfigure}{0.65\columnwidth}
         \centering  
         \includegraphics[width=\textwidth]{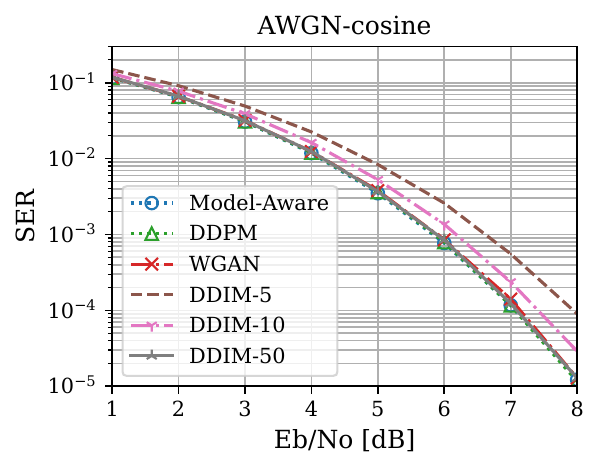}
         \subcaption{Cosine-scheduled $\beta_t$ and $v$ prediction.}\label{subfig:SER-AWGN-cosine}
    \end{subfigure}   
    \centering
    \begin{subfigure}{0.65\columnwidth}
        \centering  
        \includegraphics[width=\textwidth]{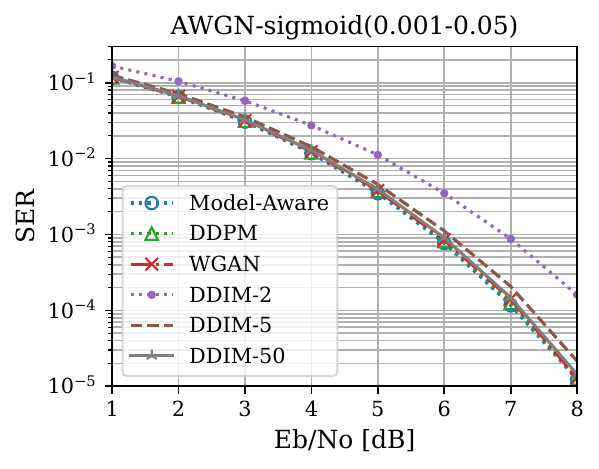}
        \subcaption{Sigmoid-scheduled $\beta_t$ with $\epsilon$ prediction.}\label{subfig:SER-AWGN-sigmoid}
    \end{subfigure}
    \centering
    \begin{subfigure}{0.65\columnwidth}
         \centering  
         \includegraphics[width=\textwidth]{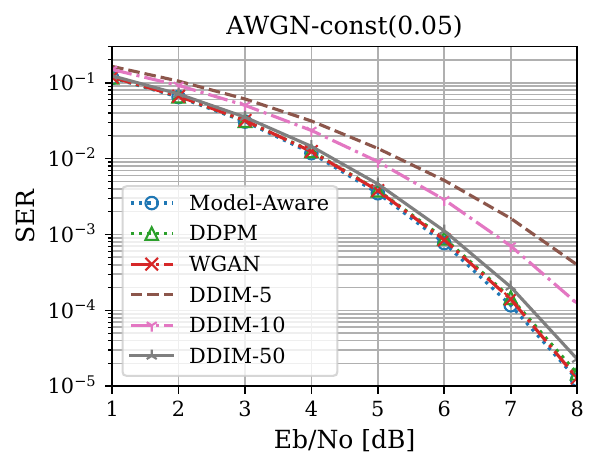}
         \subcaption{Constant $\beta_t=0.05$ with $\epsilon$ prediction.}\label{subfig:SER-AWGN-constant}
    \end{subfigure}
    \centering
    \begin{subfigure}{0.65\columnwidth}
         \centering  
         \includegraphics[width=\textwidth]{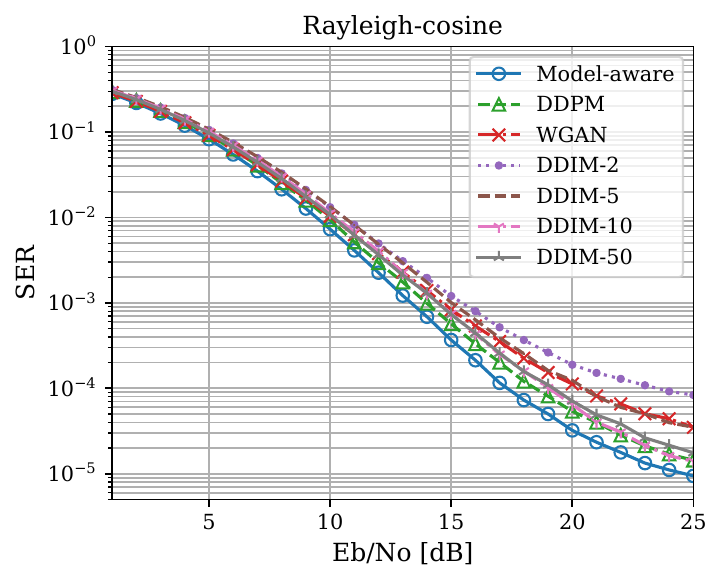}
         \subcaption{Cosine scheduled $\beta_t$ with $v$ prediction.}\label{subfig:SER-Rayleigh-cosine}
    \end{subfigure}
    \begin{subfigure}{0.65\columnwidth}
         \centering  
         \includegraphics[width=\textwidth]{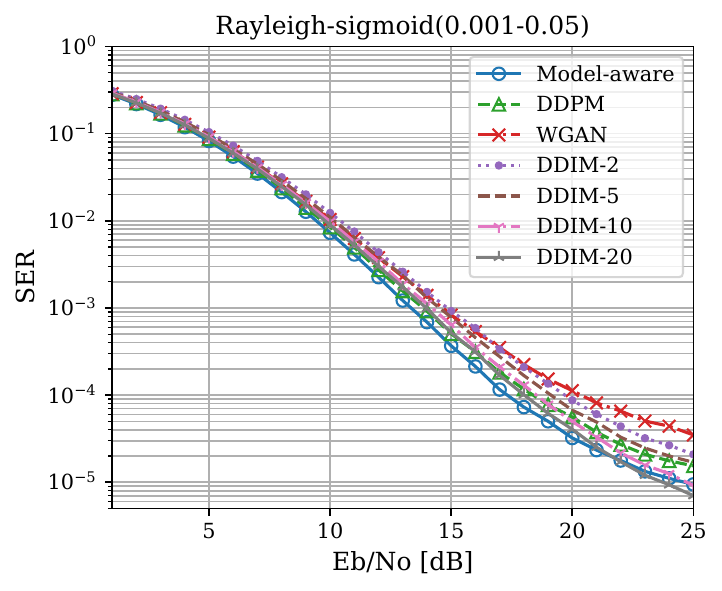}
         \subcaption{Sigmoid scheduled $\beta_t$ with $\epsilon$ prediction.}\label{subfig:SER-Rayleigh-sigmoid}
    \end{subfigure}
    \begin{subfigure}{0.65\columnwidth}
         \centering  
         \includegraphics[width=\textwidth]{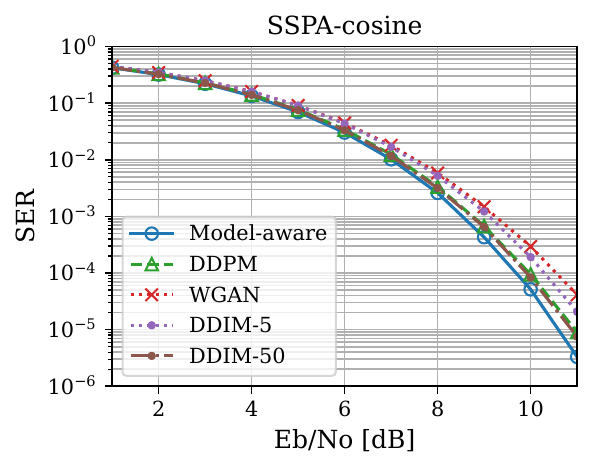}
         \subcaption{Cosine-scheduled $\beta_t$ with $v$ prediction.}\label{subfig:SER-SSPA-cosine}
    \end{subfigure}
    \caption{The E2E SER is evaluated for different sampling algorithms of DMs, i.e., DDPM and DDIM.} \label{fig:testSER_sampling}
\end{figure*}
\subsection{Sampling Acceleration by Skipped Sampling}\label{ssec:sampling_acc}
The DM's sampling can be accelerated by the skipped sampling outlined in Section~\ref{ssec:sampling_alg}. This subsection examines the trade-off between the sampling speed and mode coverage for various values of $S$ by using the SWD and E2E SER.

Sampling speed is measured for the AWGN channel by the time taken to generate $2^{17}$ samples with one NVIDIA RTX 2060 GPU.
The full sampling time of DDPM is 534~\si{\milli\second}, whereas DDIM's sampling time was measured as 4.96, 10.5, 26.9, 52.2, 98.4, 246, 492~\si{\milli\second} for $S=1, 2, 5, 10, 20, 50, 100$, respectively, and it takes 1.40~\si{\milli\second} for the WGAN. 
The sampling time increases linearly with $S$ and aligns with the DDPM's sampling time when $S$ is equal to $T$. This is due to the repetition of the denoising process for $S$ times. Notably, even when $S=100$, DDIM is faster than DDPM, presumably because the denoising process in DDPM involves random sampling of $\bm{\epsilon}$, whereas DDIM's denoising process is deterministic. 
Even though the DDPM is slower than WGAN, it can be mitigated drastically by using skipped sampling.

To evaluate DDIMs under skipped sampling, we measured the SWD with the ground truth distribution and the E2E SER when AEs are trained with them.
The DMs' SWDs are tested for the 3 channel models with constant, sigmoid, cosine noise scheduling methods and illustrated in Table \ref{Tbl:SWD}.
The SWDs were measured for $p=1$ by $128$ randomly generated projections and $10^7$ samples. The samples are generated by a normally distributed channel input and its corresponding channel output. The length $S$ of the DDIM's trajectory is denoted as DDIM-$S$ in the table, and then compared with the DDPM and a WGAN. As the skipping step size becomes larger, the SWD is observed to increase in general.

Next, the impact of skipped sampling on the E2E SER is tested and illustrated in Fig.~\ref{fig:testSER_sampling}. The E2E framework for each case was trained with the pre-training algorithm: one DDPM model was trained, and AEs were trained with samples drawn by using the DDPM sampling algorithm and the DDIM's sampling algorithms with skipped sampling. 
In Fig.~\ref{subfig:SER-AWGN-cosine}, the lowest SER values for all $E_b/N_0$ are achieved by DDPM and DDIM with $S=50$, with DDIM's SER increasing for smaller $S$ values. As a lower $S$ value reduces the sampling time but increases the SER, this constitutes a trade-off. Specifically, the sampling time, which significantly influences the training time of the AE, and the E2E communication quality are the factors involved in this trade-off. The same trend is observed for the other channel models and other noise scheduling methods in Fig.~\ref{fig:testSER_sampling} with different amount of changes. This difference made by the noise scheduling is explored in the next section.

\subsection{Diffusion Noise Scheduling}\label{ssec:beta_scheduling}
To scrutinize the influence of $\beta_t$, we evaluated three $\beta_t$-scheduling schemes, constant, cosine and sigmoid scheduling. We assessed them by the SWD and the E2E SER. The cosine and the sigmoid scheduling methods are defined as explained in Section~\ref{ssec:AWGN_channel} and in Section~\ref{ssec:ch_gen_test}, respectively. For constant and sigmoid scheduled noise, we used $\epsilon$ prediction, whereas $v$ prediction is used for cosine scheduling.   
Table \ref{Tbl:SWD} shows that the choice of noise scheduling has a significant impact on the generative performance. The cosine scheduling provides the smallest SWD for all 3 channel models and robustness to DDIM and the skipped sampling. The outstanding result of cosine scheduled noise is well aligned with the observation in \cite{ kingma2021variational, salimans2022progressive}, in that it was proposed to improve the likelihood, and the $v$ prediction is designed to be more stable for skipped sampling. The result stresses that the optimization of noise scheduling is important to improve the generative performance. 
Compared to the optimized WGAN, DMs yield smaller SWDs only for some scheduling methods and for certain level of skipped sampling. 
This highlights the need for a careful design of DMs, even though DMs are generally known to provide better mode coverage compared to GANs. 

To observe the effective influence of the noise scheduling on E2E SER, we trained AEs with various DMs and compare them in Fig.~\ref{subfig:SER-AWGN-cosine} to \ref{subfig:SER-AWGN-constant} for the AWGN channel. To reduce redundant curves, we only plotted a few curves with $S$ values that are almost optimal or starting to degrade. Compared to the constant $\beta_t$, both scheduling methods improve the test SER for all sampling algorithms, with the performance gap more pronounced for DDIMs with small $S$. In contrast to the SWDs in Table \ref{Tbl:SWD}, the sigmoid scheduling shows more robustness to DDIM sampling, resulting in a negligible SER degradation even for $S=5$. 
The same tendency is observed for the Rayleigh fading channel, plotted in Fig. \ref{subfig:SER-Rayleigh-cosine} and \ref{subfig:SER-Rayleigh-sigmoid}. Even DDIM-20 in Fig.~\ref{subfig:SER-Rayleigh-sigmoid} provides a lower SER value than the DDPM and the model-aware curve for high $E_b/N_0$ values. Remark that optimizing $\beta_t$ scheduling can be substantial for robustness to skipped sampling and accordingly for mitigating the trade-off between sampling time and generative performance.

The SWDs and the SER do not seem to evaluate the generative models with the same standards.
This is likely because some suboptimal learned distributions can be noisier than the ground truth and make the AE more robust to channel noise. 
Furthermore, the SWDs are measured for normally distributed channel inputs and averaged out over them, so the SWD indicates the ability to capture the distribution in general, with larger weighting on smaller channel input values. 
By contrast, for training AEs, a much smaller set of channel inputs is used, and the generative performance only for those channel inputs is influential. Hence, an experimental search for the optimal hyperparameters is inevitable to secure the E2E communication reliability. This could be automatized by learning the optimal noise scheduling with extra computational costs, as in \cite{kingma2021variational}.

\begin{figure*}[t!]
\centering
\begin{subfigure}{0.49\columnwidth}
    \centering
    \includegraphics[width =\columnwidth]{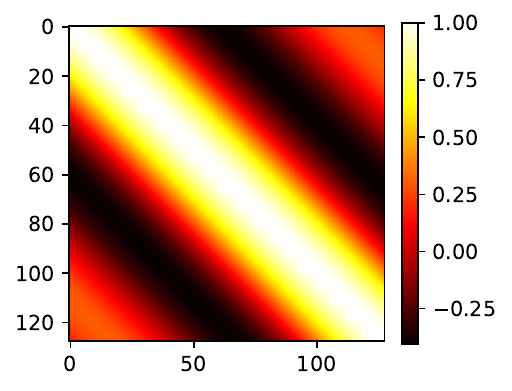}
    \subcaption{The ground truth.}\label{fig_covariance_gt}
\end{subfigure}
\begin{subfigure}{0.49\columnwidth}
    \centering
    \includegraphics[width =\columnwidth]{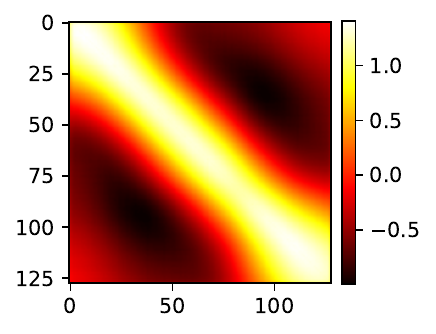}
    \subcaption{DDPM by a 1D U-Net}
\end{subfigure}
\begin{subfigure}{0.49\columnwidth}
    \centering
    \includegraphics[width =\columnwidth]{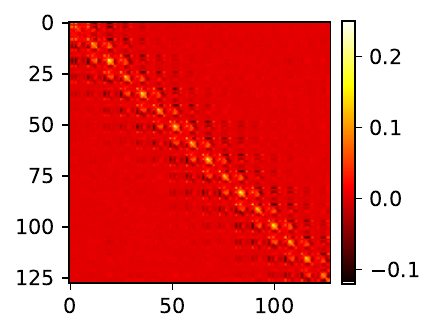}
    \subcaption{WGAN by a 1D DCGAN}
\end{subfigure}
\begin{subfigure}{0.49\columnwidth}
    \centering
    \includegraphics[width =\columnwidth]{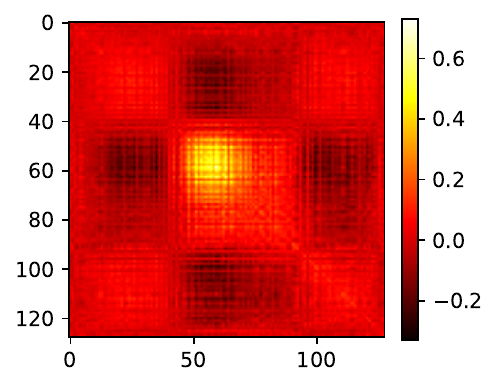}
    \subcaption{WGAN by a 1D CNN}
\end{subfigure}
\caption{The empirical covariance matrices of a Clarke's channel model that are obtained by a DM and WGANs are compared.}\label{fig_covariances}
\end{figure*}

\section{Extension to Correlated Fading Channel}
The channel models above assume channel coefficients that are independent of each other. This section tackles a channel distribution with a correlated channel matrix by using Clarke's model. 
Specifically, the channel model is identical to a complex Rayleigh fading channel model except that the elements in $\bm{H}^{n_c}$ are correlated, i.e., $\bm{Y}^{n_c}=\bm{H}^{n_c}\circ \bm{X}^{n_c}+\bm{Z}^{n_c}$, where $\bm{H}^{n_c}\sim\mathcal{CN}(\mathbf{0},\mathbf{\Sigma})$. 
The covariance matrix $\mathbf{\Sigma}$ is defined as 
\begin{align}
\mathbf{\Sigma}
&= \left[ {\begin{array}{*{20}{c}}
1 & R_h(1) & \cdots & R_h(N\!-\!1)\\
R_h(1) & 1 & \ddots & \vdots \\
 \vdots & \ddots & \ddots & R_h(1)\\
R_h(N\!-\!1) & \cdots & R_h(1) & 1
\end{array}} \right].
\label{covariance_com}
\end{align}
by the autocorrelation function $R(\ell)= J_0\left( 2\pi f_D (\ell T_s) \right)$, where $\ell$ is the number of time instances between two coefficients, $f_D=(\nu/c)f_0$ is the Doppler frequency, $\nu$ is the relative speed of the receiver, $c$ is the speed of light, $f_0$ is the carrier frequency, $\ell T_s$ is the channel symbol duration taking for $\ell$ time instants, and $J_0(\cdot)$ is the zeroth-order Bessel function, i.e., $J_0(x)=\sum_{m=0}^\infty \frac{(-1)^m}{m!m!}\left(\frac{x}{2}\right)^{2m}.$
We simulated a channel with $n=128$, $f_DT_s=0.01$, whose covariance matrix is illustrated in Fig.~\ref{fig_covariance_gt}. As the focus is the ability to generate the correlated multivariate distribution, the generative model is evaluated by the empirical covariance and the SWD. The generated channel coefficients are obtained by feeding the channel input $\bm{X}^{n_c}=(1,1,\dots, 1)$ to the DM and getting $\bm{Y}^{n_c}=\bm{H}^{n_c}+\bm{Z}^{n_c}$, and the covariance of $\bm{H}^{n_c}$ is obtained by taking $\text{Cov}(\bm{Y}^{n_c})-\sigma^2\bm{I}$. The empirical covariance is obtained by $10^3$ samples for a DM with a 1D U-Net \cite{ronneberger2015u} and WGANs based on a 1D DCGAN \cite{radford2015unsupervised} and a 1D CNN. We refer to the \texttt{GitHub} repository\footnote{\url{https://github.com/muahkim/DM_for_learning_channels}} for more details about the simulation setting to save space. The empirical covariance matrices are compared with the ground truth in Fig.~\ref{fig_covariances}. For the generated $\bm{Y}^{n_c}$ samples, the SWD is measured to be 0.25 for the DM and 0.81 and 0.89 for the WGANs with a DCGAN and a CNN, respectively, measured by using 128 projections and $10^5$ samples. Both the empirical covariance and the SWD demonstrate that the DM with a U-Net captures the covariance of the data samples better, while the WGANs fail to learn the covariance at all. This emphasizes the performance gap in learning correlation within samples, which could be substantial for generating channels when the channel is not simple enough to learn by GANs.  

\section{Conclusion and Outlook}
We examined the DDPM for generating channels within a deep learning-based E2E communication framework. For efficient training of the framework, the pre-training algorithm is proposed and evaluated.
The DMs were tested for channel models with additive random noise, random fading, non-linear power amplification, and correlated fading. 
Results showed that a DDPM could emulate the channel distributions accurately and lead to almost the same E2E SER as the model-aware case. The suggested pre-training algorithm yielded improvement in E2E SER, compared to the iterative training algorithm. While the sampling of DMs is slower than WGANs, the sampling speed could be drastically improved by using skipped sampling with DDIMs. A good choice of noise scheduling was shown to make the DDIM robust to the skipped sampling.

Upon this work as a baseline, extensions to many directions are possible, for example, to challenging channel models such as a tapped delay line model and an optical fiber model and to actual channels. For such complicated channels, the DM can be further optimized with learning-based optimization of the noise scheduling, weighting in loss, trajectory, etc. For practicality, a scalable DM and bit-wise AE could be considered for the E2E learning with a larger block length. Furthermore, a robust training algorithm of the framework to various $E_b/N_0$ values could be considered. 


{\appendix[Sampling Algorithms for $v$ Prediction]\label{appdx_sampling_v_prediction}

The DDPM denoising formula for $v$ prediction can be obtained by parameterizing $\bm{\mu}_\theta(\bm{x}_t, t)$ in terms of $\bm{x}_t$ and $\hat{\bm{v}}_
\theta(\bm{x}_t,t)$ in the following equation:
\begin{align}
\bm{x}_{t-1}(\bm{x}_t,t) &= \bm{\mu}_\theta(\bm{x}_t, t)\! + \!\sigma_t\bm{\epsilon}_t.\label{eq_ddpm_v0}
\end{align}
Equation \eqref{eq_mu_theta} of $
\bm{\mu}_\theta(\bm{x}_t,t)$ can be solved by substituting $\hat{\bm{x}}_\theta(\bm{x}_t,t)$ by $\hat{\bm{x}}_\theta(\bm{x}_t,t) = \sqrt{\bar{\alpha}_t} \bm{x}_t - \sqrt{1-\bar{\alpha}_t}\hat{\bm{v}}_\theta(\bm{x}_t,t)$ as
\begin{align}
\bm{\mu}_\theta(\bm{x}_t, t) &\!=\! \frac{\sqrt{\alpha_t}(1\!-\!\bar{\alpha}_{t-1})}{1-\bar{\alpha}_t}\bm{x}_t \!+\! \frac{\sqrt{\bar{\alpha}_{t-1}}(1\!-\!\alpha_t) }{1-\bar{\alpha}_t} \sqrt{\bar{\alpha}_t} \bm{x}_t \\
&\quad \quad - \frac{\sqrt{\bar{\alpha}_{t-1}}(1-\alpha_t) }{1-\bar{\alpha}_t} \sqrt{1-\bar{\alpha}_t}\hat{\bm{v}}_\theta(\bm{x}_t,t)\\
&=\frac{\sqrt{\alpha_t}}{1-\bar{\alpha}_t}\left(1-\bar{\alpha}_{t-1}+\bar{\alpha}_{t-1}(1-\alpha_t)\right)\bm{x}_t\\
& \quad\quad \quad - \frac{\sqrt{\bar{\alpha}_{t-1}}(1-\alpha_t) }{\sqrt{1-\bar{\alpha}_t}} \hat{\bm{v}}_\theta(\bm{x}_t,t)
\end{align}
\begin{align}
& = \sqrt{\alpha_t}\bm{x}_t - \frac{\sqrt{\bar{\alpha}_{t-1}}(1-\alpha_t) }{\sqrt{1-\bar{\alpha}_t}} \hat{\bm{v}}_\theta(\bm{x}_t,t).
\end{align}
The denoising step equation \eqref{eq_ddpm_v} can be completed by plugging in this formula of $\bm{\mu}_\theta(\bm{x}_t, t)$ in \eqref{eq_ddpm_v0}. 

The DDIM sampling algorithm with $\epsilon$ prediction is 
\begin{align}
\bm{x}_{\tau_{i-1}}&=\sqrt{\bar{\alpha}_{\tau_{i-1}}}\hat{\bm{x}}_\theta(\bm{x}_{\tau_{i}},\tau_i) +\sqrt{1-\bar{\alpha}_{\tau_{i-1}}}\hat{\bm{\epsilon}}_\theta(\bm{x}_{\tau_i},\tau_i),\label{eq_ddim_v}
\end{align}
where $\hat{\bm{x}}_\theta$ is calculated by $\bm{x}_{\tau_i}$ based on the diffusion process as in \eqref{eq_epsilon_DDIM}. 
By definition of $\bm{v}$, we have 
\begin{equation}
    \hat{\bm{x}}_\theta(\bm{x}_t,t) =\sqrt{\bar{\alpha}_t}\bm{x}_t - \sqrt{1-\bar{\alpha}_t}\hat{\bm{v}}_\theta(\bm{x}_t, t) \label{eq_xhat}
\end{equation} for all $t$. 
One can induce a similar formula for $\hat{\bm{\epsilon}}_\theta$ as a function of $\bm{x}_t$ and $\hat{\bm{v}}_\theta(\bm{x}_t,t)$, starting from the definition \eqref{eq_epsilon_theta}:
\begin{align}
\hat{\bm{\epsilon}}_\theta(\bm{x}_t, t)&=\frac{1}{\sqrt{1-\bar{\alpha}_{t}}}\left(\bm{x}_t - \sqrt{\bar{\alpha}_t} \hat{\bm{x}}_{\theta}(\bm{x}_t)\right).
\end{align}
By plugging in \eqref{eq_xhat} to the above equation, we have
\begin{align}
&\hat{\bm{\epsilon}}_\theta(\bm{x}_t, t)\\
=&\frac{1}{\sqrt{1-\bar{\alpha}_{t}}}\left(\bm{x}_t - \sqrt{\bar{\alpha}_t} \left( \sqrt{\bar{\alpha}_t}\bm{x}_t -  \sqrt{1-\bar{\alpha}_t} \hat{\bm{v}}_\theta(\bm{x}_t,t) \right)\right)\\
=&\frac{1-\bar{\alpha}_t}{\sqrt{1-\bar{\alpha}_{t}}}\bm{x}_t + \sqrt{\bar{\alpha}_t}\hat{\bm{v}}_\theta(\bm{x}_t,t) \\
=&\sqrt{1-\bar{\alpha}_{t}}\bm{x}_t + \sqrt{\bar{\alpha}_t}\hat{\bm{v}}_\theta(\bm{x}_t,t). \label{eq_epsilon_hat}
\end{align}

One can substitute $\hat{\bm{x}}_\theta(\bm{x}_{\tau_i},{\tau_i})$ and $\hat{\bm{\epsilon}}_\theta(\bm{x}_{\tau_i},{\tau_i})$ in \eqref{eq_ddim_v} by using two equations \eqref{eq_xhat} and \eqref{eq_epsilon_hat} for $t=\tau_i$, which proves the denoising formula \eqref{eq_ddim_v} as follows
\begin{align}
&\bm{x}_{\tau_{i-1}}=\sqrt{\bar{\alpha}_{\tau_{i-1}}} \left(\sqrt{\bar{\alpha}_t}\bm{x}_{\tau_i} - \sqrt{1-\bar{\alpha}_{\tau_i}}\hat{\bm{v}}_\theta(\bm{x}_{\tau_i}, \tau_i) \right) \\
 &+\!\sqrt{1\!-\!\bar{\alpha}_{\tau_{i-1}}} \left(\!\sqrt{1\!-\!\bar{\alpha}_{\tau_i}}\bm{x}_{\tau_i} \! + \! \sqrt{\bar{\alpha}_{\tau_i}}\hat{\bm{v}}_\theta(\bm{x}_{\tau_i},\tau_i)\right)\!\!\\
=& \left(\sqrt{\bar{\alpha}_{\tau_{i-1}}\bar{\alpha}_{\tau_{i}}} + \sqrt{(1-\bar{\alpha}_{\tau_{i-1}})(1-\bar{\alpha}_{\tau_{i}})} \right)\bm{x}_{\tau_i} \\
& +\!\left(\!\sqrt{\bar{\alpha}_{\tau_{i}}(1-\bar{\alpha}_{\tau_{i\!-\!1}})} \!-\!\! \sqrt{\bar{\alpha}_{\tau_{i-1}}(1-\bar{\alpha}_{\tau_{i}})} \right)\hat{\bm{v}}_\theta(\bm{x}_{\tau_i},\tau_i).\!\!
\end{align}
}

\bibliographystyle{IEEEbib}
\bibliography{refs}


\end{document}